\documentclass{aastex63}
\usepackage{CJKutf8}
\usepackage{amsmath,amsfonts}

\received{}
\revised{}
\accepted{}
\submitjournal{ApJ}

\begin{document}

\title{Characterizing the 3D Kinematics of Young Stars in the Radcliffe Wave}

\correspondingauthor{Catherine Zucker}
\email{czucker@stsci.edu}

\author[0000-0001-6099-3268]{Alan J. Tu}
\affiliation{Center for Astrophysics $\mid$ Harvard \& Smithsonian, 
60 Garden St., Cambridge, MA, USA 02138}

\author[0000-0002-2250-730X]{Catherine Zucker}
\altaffiliation{Hubble Fellow}
\affiliation{Center for Astrophysics $\mid$ Harvard \& Smithsonian, 
60 Garden St., Cambridge, MA, USA 02138}
\affiliation{Space Telescope Science Institute, 3700 San Martin Drive, Baltimore, MD 21218, USA}

\author[0000-0003-2573-9832]{Joshua S. Speagle (\begin{CJK*}{UTF8}{bsmi}沈佳士\ignorespacesafterend\end{CJK*})}
\altaffiliation{Banting \& Dunlap Fellow}
\affiliation{Department of Statistical Sciences, University of Toronto, Toronto, ON M5S 3G3, Canada}
\affiliation{David A. Dunlap Department of Astronomy \& Astrophysics, University of Toronto, Toronto, ON M5S 3H4, Canada}
\affiliation{Dunlap Institute for Astronomy \& Astrophysics, University of Toronto, Toronto, ON M5S 3H4, Canada}
\affiliation{Center for Astrophysics $\mid$ Harvard \& Smithsonian, 
60 Garden St., Cambridge, MA, USA 02138}

\author[0000-0002-8658-1453]{Angus Beane}
\affiliation{Center for Astrophysics $\mid$ Harvard \& Smithsonian, 
60 Garden St., Cambridge, MA, USA 02138}

\author[0000-0003-1312-0477]{Alyssa Goodman}
\affiliation{Center for Astrophysics $\mid$ Harvard \& Smithsonian, 
60 Garden St., Cambridge, MA, USA 02138}

\author[0000-0002-4355-0921]{Jo\~{a}o Alves}
\affiliation{University of Vienna, Department of Astrophysics, T{\"u}rkenschanzstra{\ss}e 17, 1180 Vienna, Austria}

\author[0000-0001-6251-0573]{Jacqueline Faherty}
\affiliation{Department of Astrophysics, American Museum of Natural History, Central Park West at 79th Street, NY 10024, USA}

\author[0000-0001-6879-9822]{Andreas Burkert}
\affiliation{University Observatory Munich (USM), Scheinerstrasse 1, 81679 Munich, Germany}
\affiliation{Max-Planck-Institut für extraterrestrische Physik (MPE), Giessenbachstr. 1, 85748 Garching, Germany}







\begin{abstract}

We present an analysis of the kinematics of the Radcliffe Wave, a 2.7-kpc-long sinusoidal band of molecular clouds in the solar neighborhood recently detected via 3D dust mapping. With \textit{Gaia} DR2 astrometry and spectroscopy, we analyze the 3D space velocities of $\sim 1500$ young stars along the Radcliffe Wave in action-angle space, using the motion of the wave's newly born stars as a proxy for its gas motion. We find that the vertical angle of young stars --- corresponding to their orbital phase perpendicular to the Galactic plane --- varies significantly as a function of position along the structure, in a pattern potentially consistent with a wave-like oscillation. This kind of oscillation is not seen in a control sample of older stars from \textit{Gaia} occupying the same volume, disfavouring formation channels caused by long-lived physical processes. We use a ``wavy midplane'' model to try to account for the trend in vertical angles seen in young stars, and find that while the best-fit parameters for the wave's spatial period and amplitude are qualitatively consistent with the existing morphology defined by 3D dust, there is no evidence for additional velocity structure. These results support more recent and/or transitory processes in the formation of the Radcliffe Wave, which would primarily affect the motion of the wave's gaseous material. Comparisons of our results with new and upcoming simulations, in conjunction with new stellar radial velocity measurements in \textit{Gaia} DR3, should allow us to further discriminate between various competing hypotheses.

\end{abstract}



\section{Introduction} \label{sec:intro}

The \textit{Radcliffe Wave} is a 2.7-kiloparsec-long band of molecular clouds recently discovered  \citep{Alves_2020} using new high-resolution 3D dust maps \citep{Zucker_2020, Zucker_Speagle_2019, Green_2019} enabled in part by the unprecedented astrometry of the \textit{Gaia} mission. With an aspect ratio of $20:1$, the structure contains $\gtrsim 3 \times 10^6 M_\odot$ of gas and includes a majority of nearby star-forming regions, such as the Orion, Perseus, Taurus, Cepheus, and Cygnus Molecular Clouds. The structure also aligns well with the densest part of the Local Arm of the Galaxy defined by masers \citep{Reid_2019}. As its name suggests, the Radcliffe Wave takes the form of a damped sinusoid with an estimated average period of about 2 kpc, extending above and below the Galactic disk with an amplitude of about 160 pc. 

While the structure is clearly wave-like in its 3D morphology, its 3D space motion -- and in particular, whether the structure oscillates perpendicular to the Galactic disk -- is poorly understood. Full six-dimensional information (three spatial dimensions and three velocity dimensions) is critical for discriminating between various hypotheses for the wave's origin, such as whether the structure formed due to internal effects \citep[e.g. Kelvin-Helmholtz Instability;][]{Fleck_2020, Matthews_2008} or external ones \citep[e.g. tidal interactions, collisions with high-velocity clouds or dwarf galaxies;][]{Bland_Hawthorn_2021}.

In addition to enabling significant resolution gains in 3D dust mapping, \textit{Gaia} also provides an opportunity to study the 3D kinematics of the Radcliffe Wave since thousands of young stars in the vicinity of the Sun have \textit{Gaia} radial velocity detections. When combined with additional constraints from \textit{Gaia} on the star's sky coordinates, proper motion, and parallax, we can fully constrain the properties of a star in six dimensions of position and velocity (i.e. 6D phase space). While we cannot probe the 3D space motion of molecular gas in the Radcliffe Wave directly, we can, as a proxy, rely on available \textit{Gaia} 3D velocity data of these young stars born in the wave to trace its gas motion. This assumption is warranted for two main reasons:
\begin{enumerate}
    \item Young stars are unlikely to have strayed far from their birth clouds, so young stars in the vicinity of the Radcliffe Wave probably originated from gas clouds in the wave (see \citealt{Grossschedl_2021} and references therein). 
    \item Young stars have similar radial velocities to the parental gas cloud in which they were born. This comparison also holds true in our regions of interest: several independent studies have confirmed that in the Orion, Perseus, and Taurus molecular clouds --- all constituents of the Radcliffe Wave --- stellar radial velocities serve as a good approximation for the gaseous radial velocities obtained from molecular line observations \citep{Galli_2019, Foster_2015, Grossschedl_2021, Hacar_2016}.
\end{enumerate}

While \textit{Gaia} provides 6D phase space information for stars with radial velocities, many works have found it advantageous to convert these quantities into 6D action-angle space coordinates \citep{Trick_2019, Beane_2018, Trick_2021,Sanders_2016}. \textit{Actions} are conserved quantities that describe a star's motion around the Galactic center; they assume a regular, bound orbit, as well as independence of motion in each coordinate. Stellar orbits are most conveniently studied in cylindrical coordinates where the three actions $J_R, J_{\phi}, J_z$ characterize a star's motion in the radial, azimuthal, and vertical directions. \citet{Binney_2008} argue that these quantities can be roughly interpreted in the following way:
\begin{itemize}
    \item $J_{\phi}$ is equivalent to the component $L_z$ of the angular momentum\footnote{$J_{\phi}$ and $L_z$ are only equivalent in a slowly-varying, axisymmetric potential. We assume a static axisymmetric potential in all relevant calculations used this work.}, which is proportional to the average radius of a star's orbit.
    \item $J_R$ describes the eccentricity of an orbit.
    \item $J_z$ describes the star's maximum displacement from the Galactic midplane.
\end{itemize}

While actions roughly describe the amplitude of a star's oscillation along its orbit in a given direction, their conjugate \textit{angles} (denoted by $\Omega_R$, $\Omega_{\phi}$, and $\Omega_z$) describe \textit{where} a star lies along this orbit (always falling between $0$ and $2\pi$). Actions and angles have consequently been used to study the dynamical properties of stars as a function of, e.g., age and metallicity \citep{Bland_Hawthorn_2019, Ness_2019} throughout the Milky Way.

In this paper, we analyze the kinematics of over a thousand pre-main sequence stars in the vicinity of the Radcliffe Wave to investigate whether the results reveal a large-scale oscillation or other systematic motion. We are primarily interested in the vertical motion of stars in the Radcliffe Wave. In action-angle space, a star's vertical position $z$ varies roughly sinusoidally as a function of time, with the phase of oscillation given by the vertical component of the angle, $\Omega_z$. As such, the $\Omega_z$ coordinate offers a natural approach to study the almost exclusively vertical undulation of the Radcliffe Wave. By comparing these results to those from older stars in the same volume, we can help to discriminate between various formation channels for the wave: while recent events and/or transient phenomena should only affect the vertical motions of younger stars, stable and long-term processes should impact the vertical motions of older stars as well.

The structure of this paper is as follows. In \S \ref{sec:data}, we present the young stellar sample, the clustering algorithm used to remove potential outliers, and the control sample consisting of older stars. In \S \ref{sec:methods}, we discuss how we convert these position-velocity data to action-angle coordinates and the ``wavy midplane'' model we use to parameterize the vertical components of stellar motion. In \S \ref{sec:results}, we summarize the results from fitting the model to data. In \S \ref{sec:discussion}, we compare results to the 3D structure of the gaseous Radcliffe Wave and discuss the implications our results have on constraining the origin of the Radcliffe Wave. We conclude in \S \ref{sec:conclusion}.

\section{Data} \label{sec:data}

Our primary datasets consist of two populations: a collection of young stars within 500 pc of the Sun, and a ``control'' sample of older stars with the same spatial selection. We describe the former in \S\ref{subsec:primary} and the latter in \S\ref{subsec:control}.

\subsection{Young Stars} \label{subsec:primary}

Our first data set comprises young stars from \citet{Zari_2018} within $d < 500$ pc of the Sun with astrometric measurements from \textit{Gaia} DR2 \citep{Gaia_2018}. \citet{Zari_2018} divide these stars into pre-main sequence (PMS) and upper main sequence (UMS) samples using a combination of astrometric and photometric cuts, including analysis of young stellar isochrones in color-magnitude space. Since younger stars more closely follow the motions of their birth clouds, we focus in this work on the younger PMS sample, which contains 43,719 stars with ages of $< 20$ Myr.

Furthermore, since radial velocity ($v_r$) measurements are necessary to determine a star's 3D motion and derive orbits, we restrict our analysis to only those stars in the PMS sample with radial velocity measurements. This cut leaves us with 4,346 stars. While ideally we would limit our analysis to only include the youngest stars ($< 5$ Myr), we found this drastically reduced the size of our sample and so opted to keep the more permissive $< 20$ Myr limit in our subsequent analysis.

\begin{figure}
\begin{center}
\includegraphics[width=1.0\columnwidth]{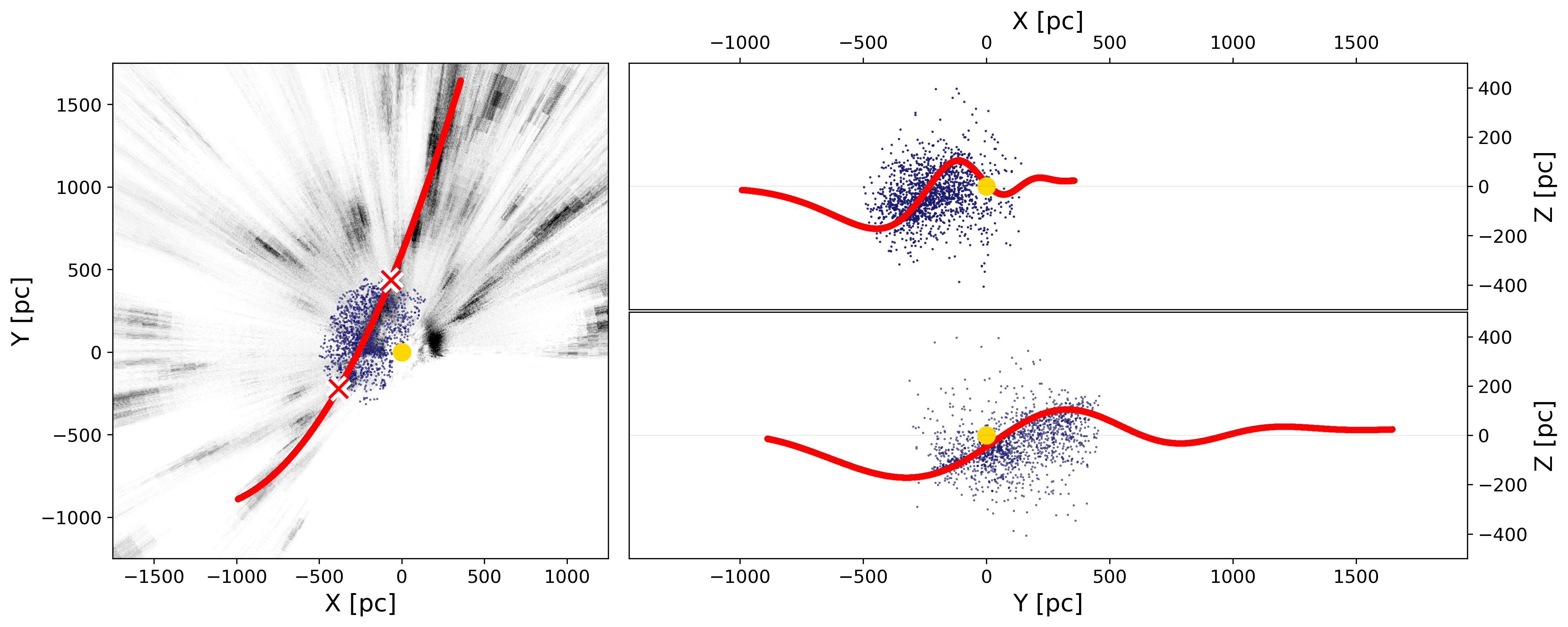} 
\caption{3D distribution of young stars alongside the the Radcliffe Wave. We show XY (left), XZ (top right), and YZ (bottom right) projections. The navy points represent the pre-main sequence stars \citep{Zari_2018} we use for our primary analysis without any clustering applied. The red line indicates the best-fit model of the Radcliffe Wave \citep{Alves_2020} detected in 3D dust. The 3D dust map from \citet{Green_2019} integrated over the range $|Z| < 300$ pc is shown in the background gray-scale in the left panel. The position of the Sun is marked with a yellow dot. The two red crosses mark the beginning $(X, Y = -385,-222)$ pc and ending $(X, Y = -67, 438)$ pc positions along the wave that we consider in Figure \ref{fig:big3}.}
\label{fig:first}
\end{center}
\end{figure}

From this sample, we then aim to select stars that lie in the vicinity of the Radcliffe Wave. We define this as any star that lies within 240 pc (i.e. twice the estimated 120 pc width) of the best-fit wave model derived from 3D dust in a top-down XY projection \citep{Alves_2020}. This is a generous cut intended to include as many potentially associated stars as possible given the relatively small size of the radial velocity sample. We do not make a cut along the $z$-axis since there is no corresponding cut made in \citet{Zari_2018}. This leaves us with a final sample of 1,522 stars, which we hereafter refer to as the \textit{unclustered Radcliffe Wave sample}. In Figure \ref{fig:first}, we show XZ, YZ, and XY projections of this sample, alongside the best-fit model of the Radcliffe Wave from \citet{Alves_2020} used in our selection. 

As explained in \citet{Zari_2018}, the PMS sample likely contains contaminating sources from the binary sequence, in particular unresolved binaries, which are unlikely to have been born in the molecular clouds defining the wave. Young stars tend to be born in clusters \citep{Lada_2003}, so we apply the Density-Based Spatial Clustering of Applications with Noise ({\tt\string DBSCAN}) clustering algorithm \citep{Ester_1996} to clean the Radcliffe Wave sample. The {\tt\string DBSCAN} algorithm finds clusters in large spatial datasets by exploring the local density of data points. Following the technique described in \citet{Chen_2020}, we keep the hyper-parameters the same ({\tt\string eps} = 0.5 and {\tt\string min\_samples} = 20) but perform the analysis in 3D position space rather than 5D position and proper motion space due to the smaller sample size. This procedure removes 383 out of the 1522 stars. We shall refer to this remaining sub-sample of 1139 stars as the \textit{clustered Radcliffe Wave sample}. We perform a subset of the model fitting using both the clustered and unclustered samples in order to gauge its effect on our results. 

As an additional check on the fidelity of the young star sample, we determine whether the sample shows evidence of variability, as given by the variability amplitude metric, $A_{var}$ defined in \citet{Deason_2017} (see their Equation 2): $A_{var} = \sqrt{N_{obs}} \times \sigma(F)/F$. The $N_{obs}$ parameter is the number of CCD crossings (we only consider the ``good" CCD crossings, $N_{G}AL$ in the \textit{Gaia} archive) and $F$ and $\sigma(F)$ are the \textit{Gaia} $G$ band flux and flux error. We find that all but two stars in the ``clustered" sample and all but six stars in the ``unclustered" sample show evidence of variability $A_{var}$ at a level greater than 1\%, validating the sample from \citet{Zari_2018}.

Finally, we assign each star an associated position ``along'' the Radcliffe Wave in order to quantify behaviour that may occur more naturally in a Radcliffe Wave-centric coordinate system (rather than an XYZ one). To do so, we divide the portion of the Radcliffe Wave model that overlaps with the \citet{Zari_2018} catalog at distances $< 500$ pc into bins (such that each bin has approximately 50 stars) and then assign stars to bins based on their projected position. In other words, we measure distance along the red curve in the left panel of Figure \ref{fig:first}, which is a projection onto the XY plane and thus independent of $z$ position. A projected position value of 0 pc corresponds to the point $(X,Y) = (-385, -222)$ pc, near the trough of the wave and the position value increases in the positive $x$ and $y$ directions. A position value close to 700 pc corresponds to the peak of the wave with coordinates $(X,Y) = (-67, 438)$ pc. See Figure \ref{fig:first} to see where these points lie in XY-space.

\subsection{Control Sample} \label{subsec:control}

To contextualize our findings based on young stars, we also construct a ``control'' sample of older stars. This sample, which consists of a random selection of \textit{Gaia} main sequence stars, is designed to ensure that any trends we uncover are restricted to the young stars, rather than all stars in the vicinity of the Radcliffe Wave. 

Before our random selection, we make cuts on the over 7 million \textit{Gaia} DR2 stars with radial velocity measurements to match the cuts made in \citet{Zari_2018}\footnote{\citet{Zari_2018} apply a series of cuts to select their young star sample, including a cut on the minimum parallax ($\rm \pi > 2 \; mas$) and on the signal to noise of the parallax measurements (\texttt{parallax\_over\_error} $>$ 5), before imposing a cut above the 20 Myr isochrone in a dereddened \textit{Gaia} CMD to ensure the sample is young. To select our control sample, we simply remove the isochrone-based cuts but keep all other astrometric cuts. Our requirement of a \textit{Gaia} $A_G$ measurement facilitates a comparison between the dereddened CMD for pre-main sequence stars shown in \citet{Zari_2018} alongside our older control sample, which we show in the Appendix.}. This includes:
\begin{itemize}
    \item Removing all stars with $d \geq 500$ pc.
    \item Removing all stars where \texttt{parallax\_over\_error} $< 5$.
    \item Removing all stars without an inferred $G$-band extinction ($A_G$).
\end{itemize}
We then perform the same spatial selection as in \S\ref{subsec:primary} to select only stars in the vicinity of the Radcliffe Wave. 

From the approximately half a million remaining stars, we finally randomly select a subset of stars to match the number of stars in the unclustered Radcliffe sample. So as not to choose too many stars in regions with very high stellar densities, we perform a stratified subsampling along the Radcliffe wave by dividing it into 100 equally-sized sectors and then sampling $\approx$ 15 stars from each. These stars are then also assigned a corresponding projected position along with Radcliffe Wave, analogous to the young stars in \S\ref{subsec:primary}. Differences between the two samples are further highlighted in Appendix \ref{sec:CMD}, which shows the dereddened CMD for the unclustered young stars alongside the main sequence control samples.

\section{Methods} \label{sec:methods}

Our overall methodology is straightforward. First, we convert our 6D phase space measurements (positions and velocities) to associated action-angle coordinates. Then, we want to use the distribution of the vertical angles $\Omega_z$ across our samples to constrain possible variations in the orbital midplane that may be associated with the Radcliffe Wave. In \S\ref{subsec:conversion}, we describe our approach to computing action-angle coordinates. In \S\ref{subsec:model}, we provide an overview of our modelling approach.

\subsection{Conversion to Action-Angle Space} \label{subsec:conversion}

Computing actions in a known potential is a numerically challenging problem, but several methods are now publicly available \citep[see e.g.][for a review of different methods]{2016MNRAS.457.2107S}. We use three different methods for computing stellar actions in this work. In decreasing order of both accuracy and computational expense they are: 
\begin{enumerate}
    \item direct orbit integration,
    \item the St\"ackel-Fudge method, and
    \item the epicyclic approximation.
\end{enumerate}
We will only use the direct orbit integration to explore general trends in vertical angle (and not for our subsequent model fitting) due to its large computational expense.

The first method we use is direct orbit integration, which we consider our ``gold standard" method. After transforming the stars in both the young and old samples to Heliocentric Galactic Cartesian coordinates using the \texttt{astropy} package \citep{astropy}, we employ the {\tt\string gala} \citep{Price-Whelan_2017} implementation of the algorithm in \citet{2014MNRAS.441.3284S} to transform the stars to action-angle coordinates. We use {\tt\string gala}'s built-in {\tt\string MilkyWayPotential}, which is based on the \texttt{MWPotential2014} in \texttt{galpy} \citep{2015ApJS..216...29B} and includes a \citet{1990ApJ...356..359H} bulge and nucleus, a Miyamoto-Nagai disk \citep{1975PASJ...27..533M}, and a Navarro-Frenk-White halo \citep{1997ApJ...490..493N}. For each orbit, we attempt to integrate with $10,000$ time steps of $0.5\,\text{Myr}$ each ($\sim5\,\text{Gyr}$, or $\sim 20$ orbits for a Sun-like star) using the Dormand-Prince integrator \citep{DORMAND198019}. If this procedure fails, we try again with 400,000 timesteps of 0.05 Myr each (after which all stars converge). As a result of this procedure, each star's 6D position and velocity are now expressed via three action coordinates ($J_R$, $J_{\phi}$, $J_z$) and three angle coordinates ($\Omega_R$, $\Omega_{\phi}$, $\Omega_z$). 

Next, we use the \texttt{galpy} \citep{2015ApJS..216...29B} implementation of the {\em St\"ackel-Fudge} method \citep{2012MNRAS.426.1324B}. This method relies on locally approximating the potential as a  St\"ackel potential \citep{1985MNRAS.216..273D}, and is commonly used as a fast but accurate approximation, especially for disk-like orbits \citep[e.g.][]{2018ApJ...856L..26M,Trick_2019,2019MNRAS.490.1026H}. For our potential we assume the standard axisymmetric and time-invariant \texttt{MWPotential2014} provided in \texttt{galpy}.

The final method we use to compute actions and angles is through the epicyclic approximation, which makes the assumption that the dynamics in the $R$ and $z$ directions are decoupled \citep[for an introduction, see \S 3.2.3 of][] {Binney_2008}. Therefore, while the previous two methods computed all three actions and angles at the same time, we can compute the vertical action and angle in isolation here. Following \citet{Beane_2019}, the vertical motion is assumed to be that of a harmonic oscillator  with:
\begin{equation} \label{eq:z}
z = A_z \sin (\nu t + \beta) + z_0
\end{equation}
\begin{equation} \label{eq:vz}
v_z \equiv \frac{{\rm d} z}{{\rm d} t} = A_z \nu \cos (\nu t + \beta) + v_{z_0}
\end{equation}
Here $A_z$ is the amplitude of the orbit, $\nu$ is the vertical restoring frequency, $\beta$ is an arbitrary phase, and $z_0$ and $v_{z_0}$ are the location and velocity of the midplane. Then, the vertical angle can be computed as
\begin{equation} \label{eq:omegaz}
\Omega_z \equiv \nu t + \beta = \tan^{-1}\left(\frac{z - z_0}{v_z - v_{z_0}} \nu \right)
\end{equation}
When using the epicyclic approximation for our subsequent modeling, we allow the value of $\nu$ to be a free parameter, although we assume there is no large-scale spatial variation. We also allow the location and velocity of the midplane to vary with position along the Radcliffe Wave. 

We use the latter two methods (the St\"ackel-Fudge method and the epicyclic approximation) in our model fitting, described in \S \ref{subsec:model}. We emphasize that these two methods are complementary and have differing sets of advantages and disadvantages. The St\"ackel-Fudge method is more accurate than the epicyclic approximation. However, in the epicyclic approximation we do not need to assume a Galactic potential by allowing the vertical restoring frequency $\nu$ (which would depend on an assumed potential) to be a free parameter in our model fitting. Therefore, we use St\"ackel-Fudge and the epicyclic approximation in tandem to build confidence in the robustness of our results.

The accuracy of the epicylic approximation (and by extension the St\"ackel-Fudge method) is confirmed in Appendix \ref{sec:epicomp} through comparison with the direct orbit integration results computed from {\tt\string gala}. The typical error is $\approx 0.2$ radians ($\approx 3\%$). Online at the \href{https://doi.org/10.7910/DVN/58Z87S}{Harvard Dataverse}, we provide a table summarizing the young star results for the unclustered Radcliffe Wave sample and refer interested readers there for more details (doi:10.7910/DVN/58Z87S). 

\subsection{Fitting a Model} \label{subsec:model}

In order to parameterize potential oscillatory behavior in action-angle space, we need to build a sinusoidal model to describe the structure of the Radcliffe Wave sample in position and velocity space. We will then compare this model to the action-angle data for each star, computed using both the epicylic and St\"ackel-Fudge approximations. 

The epicyclic approximation \citep{Binney_2008}, which is quite accurate for a thin disk orbit, rests on the assumption that the orbit of a given star follows simple harmonic motion independently in the $R$ and $z$ directions. Following the epicyclic approximation from \S\ref{subsec:conversion}, Equation \ref{eq:omegaz} provides a parameterization of the phase of the orbit in the $z$ direction $\Omega_z$, as a function of the vertical restoring frequency $\nu$ and the location ($z_0$) and velocity ($v_{z_0}$) of the midplane. Rather than setting $z_0$ and $v_{z_0}$ to be constants, we instead assume that they may be spatially varying as a function of position $s$ along the Radcliffe Wave in XY space (see \S\ref{sec:data}) according to the following functional forms:
\begin{equation} \label{eq:z0}
z_0(s) = B \sin (\omega s + \gamma) + C
\end{equation}
\begin{equation} \label{eq:vz0}
v_{z_0}(s) = D \cos (\omega s + \gamma) + E
\end{equation}
where $B$ represents the amplitude of the positional function, $C$ represents the offset of the positional function, $D$ represents the amplitude of the velocity function, $E$ represents the offset of the velocity function, $\omega$ represents angular frequency, and $\gamma$ represents a phase shift. In other words, we allow for the dynamical midplane to be ``wavy'' rather than flat.

\begin{figure}
\begin{center}
\includegraphics[width=1.0\columnwidth]{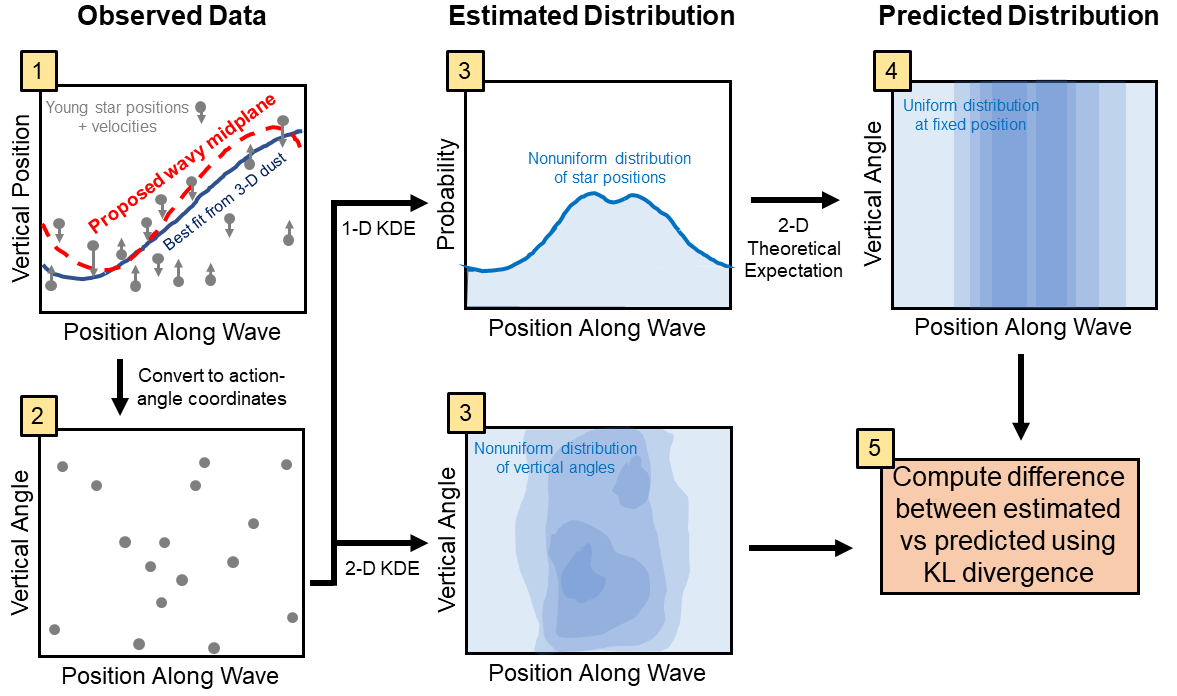} 
\caption{A schematic illustration of our modelling approach. For a proposed wavy midplane model, we take our sample of young stars (selected relative to the best-fit model of the Radcliffe Wave estimated from 3D dust) shown in the top left panel and convert their corresponding 6D positions and velocities to their corresponding action-angle coordinates. This then provides a corresponding distribution of vertical angles, as shown in the bottom left panel. Using kernel density estimation (KDE), we then estimate both the 1-D distribution in projected position along the Radcliffe Wave as well as the 2-D distribution in both projected position and vertical angle (middle panels). Based on the estimated 1-D distribution of projected positions, we then compute the expected 2-D distribution assuming all stars orbit around the proposed midplane (top right panel). We then compare the two distributions using a scaled version of the Kullblack-Leibler (KL) divergence (bottom right panel). See \S\ref{subsec:model} for additional details.}
\label{fig:schematic}
\end{center}
\end{figure}

\begin{deluxetable*}{clcc}
\tablenum{1}
\tablecaption{Priors for the Wavy Midplane Model} \label{tab:priors}
\tablewidth{0pt}
\tablehead{
\colhead{Parameter} & \colhead{Description} & \colhead{Prior} & Units 
}
\decimalcolnumbers
\startdata
$B$ & Midplane position amplitude & ${\rm Uniform}(0, 300)$ & pc \\
$\omega$ & Angular frequency & ${\rm Uniform}(0, 4\pi)$ & kpc$^{-1}$  \\
$\gamma$ & Phase shift & ${\rm Uniform}(0, 2\pi)$ & rad  \\
$C$ & Midplane position offset & ${\rm Uniform}(-200, 200)$ & pc \\
$D$ & Midplane velocity amplitude & ${\rm Uniform}(-15, 15)$ & km s$^{-1}$ \\
$E$ & Midplane velocity offset & ${\rm Uniform}(-15, 15)$ & km s$^{-1}$ \\
$\nu$ & Epicyclic frequency & ${\rm Uniform}(0, 200)$ & km s$^{-1}$ kpc$^{-1}$ \\
\enddata
\tablecomments{Choice of priors for the model fits described in \S \ref{subsec:model}, which we assume to be uniform over the listed ranges. Note that the epicyclic frequency $\nu$ is only modeled when using the epicyclic approximation; it is not used with the St\"ackel-Fudge approximation, which assumes a known Milky Way potential.}
\end{deluxetable*}

Assuming the epicylic approximation holds, our resulting set of $\{\Omega_{z,i}(\theta)\}_{i=1}^{n} \equiv \{ \Omega_{z,1}(\theta), \dots, \Omega_{z,n}(\theta)\}$ values (one for each star $i$ in our sample of $n$ stars) given the parameters $\theta \equiv (B, \omega, \gamma, C, D, E, \nu)$ that correctly describe the dynamical midplane should be \textit{uniformly distributed in angle at fixed position along the wave $s$}.\footnote{Such a situation would arise if the self-gravity of the Radcliffe Wave is important, so that stars follow an orbit centered on the Radcliffe wave instead of the midplane of the Galactic disk. Given the mass and height of the Radcliffe wave, we do expect its self-gravity to be important.} To determine the degree to which this behaviour holds, we therefore want to compare our observed probability distribution
\begin{equation}
    P_{\rm obs}(\Omega_z, s | \theta) = P_{\rm obs}(\Omega_z | s, \theta) \times P_{\rm obs}(s)
\end{equation}
to our theoretical prediction that the conditional distribution is uniform, i.e.
\begin{equation}
    P_{\rm unif}(\Omega_z, s | \theta) = \frac{1}{2\pi} \times P_{\rm obs}(s) \quad \Longleftrightarrow \quad P_{\rm obs}(\Omega_z | s, \theta) = \frac{1}{2\pi}
\end{equation}
where the conditional dependence on $\theta$ arises from the fact that changing $\theta$ affects the distribution of vertical angle ($\Omega_{z}$) but not projected position along the Radcliffe Wave ($s$).

For a given set of parameters $\theta$, we quantify the non-uniformity in the distribution by defining a (pseudo-)log-likelihood $\ln \mathcal{L}_{\rm KL}(\theta)$ based on the \textit{Kullback-Leibler (KL) divergence} between $P_{\rm obs}(\Omega_z, s | \theta)$ and $P_{\rm unif}(\Omega_z, s | \theta)$:
\begin{align} \label{eq:kl}
\ln \mathcal{L}_{\rm KL}(\theta) &\equiv - \xi \int P_{\rm unif}(\Omega_z, s | \theta) \, \left[ \ln P_{\rm unif}(\Omega_z, s | \theta) - \ln P_{\rm obs}(\Omega_z, s | \theta) \right] \,{\rm d}\Omega_z {\rm d}s \\
&= \frac{\xi}{2\pi} \int P_{\rm obs}(s) \, \ln P_{\rm obs}(\Omega_z | s, \theta) \, {\rm d}\Omega_z {\rm d}s \: + \: {\rm constant}
\end{align}
where $\xi$ is a scaling constant to help ensure appropriate uncertainty estimates.
We can interpret this likelihood as a weighted average over the difference in the log-probabilities between $P_{\rm unif}(\Omega_z, s | \theta)$ (based on our hypothetical, truly uniform sample) and $P_{\rm obs}(\Omega_z, s | \theta)$ (based on the Radcliffe Wave sample) where the weight as a function of $\Omega_z$ and $s$ is specified by the expected theoretical distribution $P_{\rm unif}(\Omega_z, s)$. This quantity is maximized when the two distributions being compared are identical, i.e. when $P_{\rm unif}(\Omega_z, s | \theta) = P_{\rm obs}(\Omega_z, s | \theta)$or, as can be seen in the simplified version, when $P_{\rm obs}(\Omega_z | s, \theta) = 1/2\pi$. The negative sign ensures that as the KL divergence gets smaller, our likelihood improves.

To evaluate our (pseudo-)log-likelihood function $\ln \mathcal{L}_{\rm KL}(\theta)$, we need to estimate $P_{\rm obs}(s)$ and $P_{\rm obs}(\Omega_z, s | \theta)$ from the set of $1,000-1,500$ stars in our sample. We use a \textit{kernel density estimation (KDE)} approach with a Gaussian kernel whose width is chosen to be a few percent of the overall allowed range in $s \in [0, 750)$ pc and $\Omega_z \in [0, 2\pi)$ rad to compute an estimate for $P_{\rm obs}(s)$ (once) and $P_{\rm obs}(\Omega_z, s | \theta)$ (each time we change $\theta$). To compute an estimate of the integral, we evaluate our KDE-based estimates for a given $\theta$ over a uniformly-spaced grid spanning the same range in $s$ and $\Omega_z$; we find a $50 \times 50$ grid is accurate enough for our purposes while remaining computationally light.

After finding the best fit value $\theta_{\rm best}$ from the data based on the unclustered young star sample, we simulate $m$ realizations of the $n$ data points using the same estimate for $P_{\rm obs}(s)$ derived previously and assuming a uniform distribution in $P(\Omega_z | s, \theta)$. We then compute the best-fit parameters $\{ \theta_{{\rm best}, j} \}_{j=1}^{m}$ for each realization. After we confirm that the best-fit solutions are unbiased relative to our simulations, we use them to determine $\xi$ so that the $1\sigma$ range in estimated values corresponds to a log-likelihood reduction of $\approx 1$, which is the expected difference assuming the distribution follows a multivariate Gaussian. We find after $m=100$ iterations that $\xi \approx 1000$ largely satisfies this constraint.

A schematic illustration of our modelling approach is shown in Figure \ref{fig:schematic}.

\begin{figure}
\begin{center}
\includegraphics[width=1.0\columnwidth]{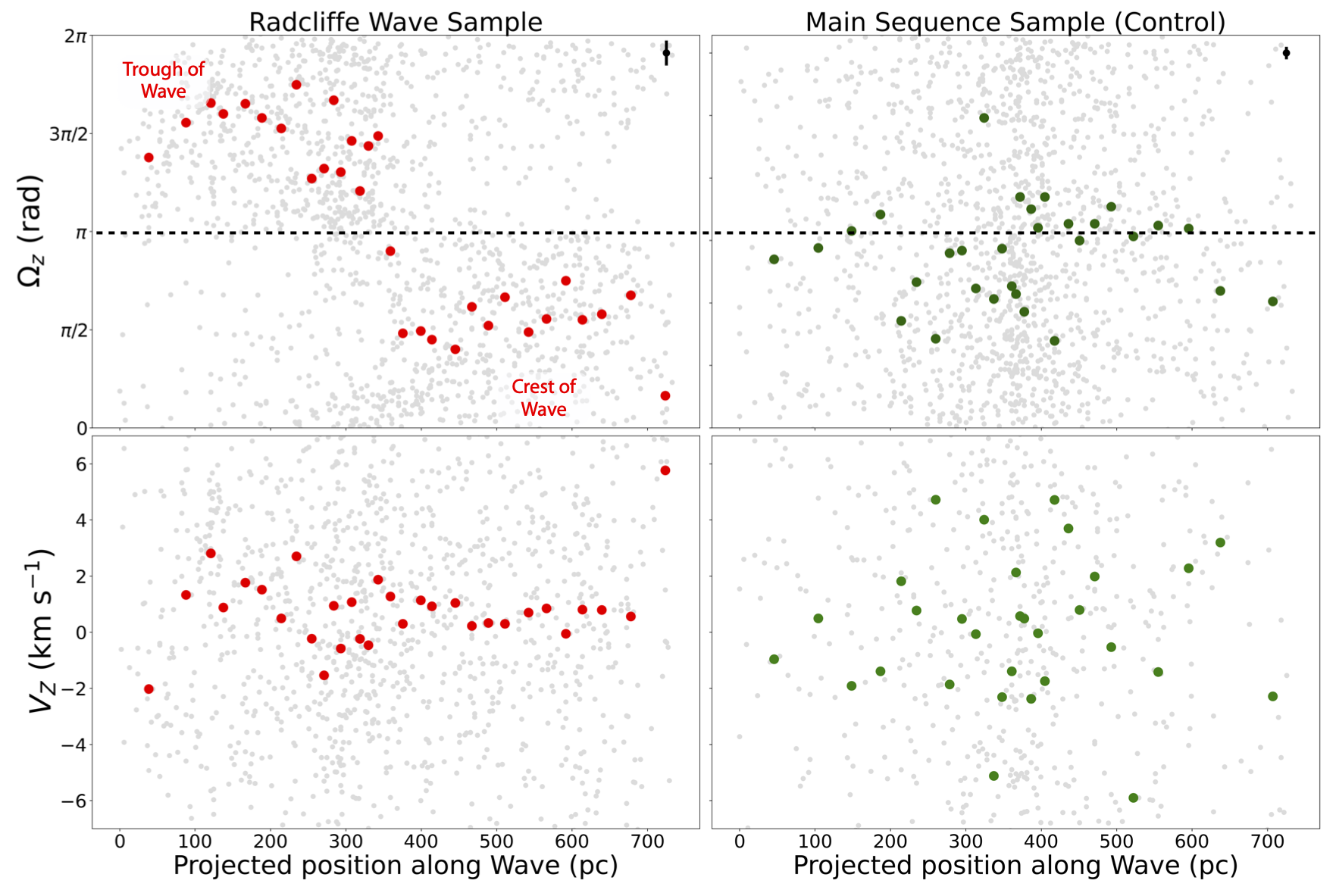} 
\caption{\textbf{Top}: Vertical angle vs. projected position along the Radcliffe Wave for the Radcliffe Wave pre-main sequence star sample (left) and the older Main Sequence control sample (right). The grey points represent the stars in each sample and the colored dots are the binned ``phase-marginalized'' medians. In the upper right hand corner of each panel we show the typical error on the vertical angle for each sample in black. In the Radcliffe Wave sample of young stars, we see a clear linear trend in phase space consistent with a wave-like oscillation. No such trend exists in the older stellar population, implying that this is not purely due to the cuts used to select both samples. \textbf{Bottom}: Same as the top panel, but for the vertical velocity $v_z$ vs. projected position along the Radcliffe Wave. Unlike with the vertical angle, we find no systematic trend in the vertical velocities of the young stars as a function of position along the wave and we detect no fluctuations in the average vertical velocity at a level $\rm \gtrapprox 3 \; km\; s^{-1}$.}
\label{fig:big3}
\end{center}
\end{figure}

To estimate uncertainties on the parameters given our observed data $P(\theta | \{ s_i \}, \{\Omega_{z_i}\})$, we use Bayes Theorem, which can be written as:
\begin{equation}
    P(\theta | \{ s_i \}_{i=1}^{n}, \{\Omega_{z_i}\}_{i=1}^{n}) \propto P(\{ s_i \}_{i=1}^{n}, \{\Omega_{z_i}\}_{i=1}^{n} | \theta) P(\theta) \equiv \mathcal{L}_{\rm KL}(\theta) \pi(\theta)
\end{equation}
where $P(\theta | \{ s_i \}_{i=1}^{n}, \{\Omega_{z_i}\}_{i=1}^{n})$ is the posterior distribution, $P(\{ s_i \}_{i=1}^{n}, \{\Omega_{z_i}\}_{i=1}^{n} | \theta) = \mathcal{L}_{\rm KL}(\theta)$ is our likelihood in Equation \eqref{eq:kl} and $P(\theta) \equiv \pi(\theta)$ is our prior. We assume flat priors on all parameters, as summarized in Table \ref{tab:priors}. 

We generate samples from the posterior distribution using the nested sampling program $\tt\string dynesty$ \citep{Speagle_2020}. We use the default configuration and stopping criteria, although we place a periodic boundary condition on the parameter $\gamma$. This fitting process is performed with both the clustered and unclustered Radcliffe Wave samples and with the epicyclic and St\"ackel-Fudge approximations. As a control test, we also perform the fit to the older, main sequence control sample with the epicyclic approximation. 

\section{Results} \label{sec:results}

\renewcommand{\arraystretch}{1.5}
\begin{deluxetable*}{ccccc}
\tablenum{2}
\tablecaption{Parameter Estimates from the Wavy Midplane Model} \label{tab:bfparams}
\tablewidth{0pt}
\tablehead{
\colhead{Parameter} & \colhead{Epicyclic (Unclustered)} & \colhead{Epicyclic (Clustered)} & \colhead{Stäckel (Unclustered)} & \colhead{Stäckel (Clustered)}
}
\decimalcolnumbers
\startdata
$B$ & $68_{-7}^{+6}$ pc & $96_{-18}^{+15}$ pc & $65_{-9}^{+167}$ pc & $79_{-6}^{+12}$ pc \\
$\omega$ & $6.1_{-0.9}^{+1.3}$ kpc$^{-1}$ & $4.6_{-0.9}^{+2.1}$ kpc$^{-1}$ & $5.5_{-4.4}^{+1.5}$ kpc$^{-1}$ & $5.3_{-1.2}^{+0.9}$ kpc$^{-1}$ \\
$\gamma$ & $4.2_{-0.4}^{+0.4}$ & $4.6_{-0.6}^{+0.4}$ & $4.5_{-0.6}^{+1.6}$ & $4.5_{-0.3}^{+0.5}$ \\
C & $-12_{-8}^{+5}$ pc & $-2_{-10}^{+10}$ pc & $-12_{-53}^{+15}$ pc & $-11_{-6}^{+5}$ pc \\
D & $0.3_{-0.9}^{+0.9}$ km s$^{-1}$ & $1.5_{-1.6}^{+1.9}$ km s$^{-1}$ & $0.2_{-2.9}^{+4.1}$ km s$^{-1}$ & $1.6_{-1.7}^{+1.7}$ km s$^{-1}$ \\
E & $0.6_{-0.7}^{+0.7}$ km s$^{-1}$ & $-0.5_{-1.6}^{+1.3}$ km s$^{-1}$ & $0.7_{-4.2}^{+2.3}$ km s$^{-1}$ & $-0.4_{-1.5}^{+0.9}$ km s$^{-1}$ \\
$\nu$ & $91_{-8}^{+9}$ km s$^{-1}$ kpc$^{-1}$ & $119_{-10}^{+11}$ km s$^{-1}$ kpc$^{-1}$ & --- & --- \\
\enddata
\tablecomments{Results of the model fitting described in \S \ref{subsec:model} for both the clustered and unclustered Radcliffe Wave samples, fit using both the epicyclic and St\"ackel-Fudge approaches. For each parameter, the median and the \textit{$2\sigma$} uncertainties (i.e. 95\% credible intervals) are reported, as derived from the \texttt{dynesty} nested sampling results. See Table \ref{tab:priors} for a summary of the various parameters and priors used in the analysis.}
\end{deluxetable*}

In the top panel of Figure \ref{fig:big3}, we plot the vertical angle $\Omega_z$ versus each star's projected position along the wave for the primary (unclustered) young star Radcliffe Wave sample and older Main Sequence control sample, where $\Omega_z$ has been computed using the ``gold standard" orbit integration with \texttt{gala} as described in \S \ref{subsec:conversion} without any modifications to the orbital midplane. The individual stars in each sample are shown in light gray. The colored dots represent the median phase of the stars in each bin. Projected position is again defined as the two-dimensional distance along the wave when considering only the $x$ and $y$ coordinates. To account for periodic uncertainties in $\Omega_z$, we compute a ``phase-marginalized'' median in each bin by perturbing individual angles based on their estimated measurement uncertainty and also allowing the entire bin to experience random shifts in angle. This helps us account for possible periodic effects that may occur if the vertical angles for some stars vary around $0$ or $2\pi$ rad.

As seen in the top left panel of Figure \ref{fig:big3}, there is a noticeable trend in the vertical angle of the young star sample as we move along the Radcliffe Wave. The stars nearer to the trough of the wave generally have $\Omega_z \approx \frac{3\pi}{2}$, whereas the the stars nearer to the peak of the wave generally have $\Omega_z  \approx \frac{\pi}{2}$. The trend in vertical angle is noticeably not present in the older Main Sequence sample, as seen in the top right panel of Figure \ref{fig:big3}. In the bottom panels of Figure \ref{fig:big3}, we plot vertical velocity as a function of projected position along the Radcliffe Wave for both the young star sample and the main sequence control sample. Unlike with the vertical angle, we observe no significant trend (at a level greater than $\approx$ $\rm 3.0 \; km \; s^{-1}$) in the vertical velocities of young stars as a function of position along the Radcliffe Wave. 
 
However, by convention, the vertical velocity $v_z$ is negative when the vertical angle is between $\frac{\pi}{2}$ and $\frac{3\pi}{2}$, while the vertical velocity is positive when the vertical angle is between $\frac{3\pi}{2}$ and $\frac{\pi}{2}$. Despite small average variations in the vertical velocity along the wave, the strong trend in vertical angle still suggests that the stars at negative $z$ (below the Galactic plane) are ``on their way up'' in their vertical motion (transitioning from negative $v_z$ to positive $v_z$) while the stars at positive $z$ (above the Galactic plane) are ``on their way down'' (transitioning from positive $v_z$ to negative $v_z$), which is \textit{fully consistent with a wave-like oscillation}.

\begin{figure}
\begin{center}
\includegraphics[width=1.0\columnwidth]{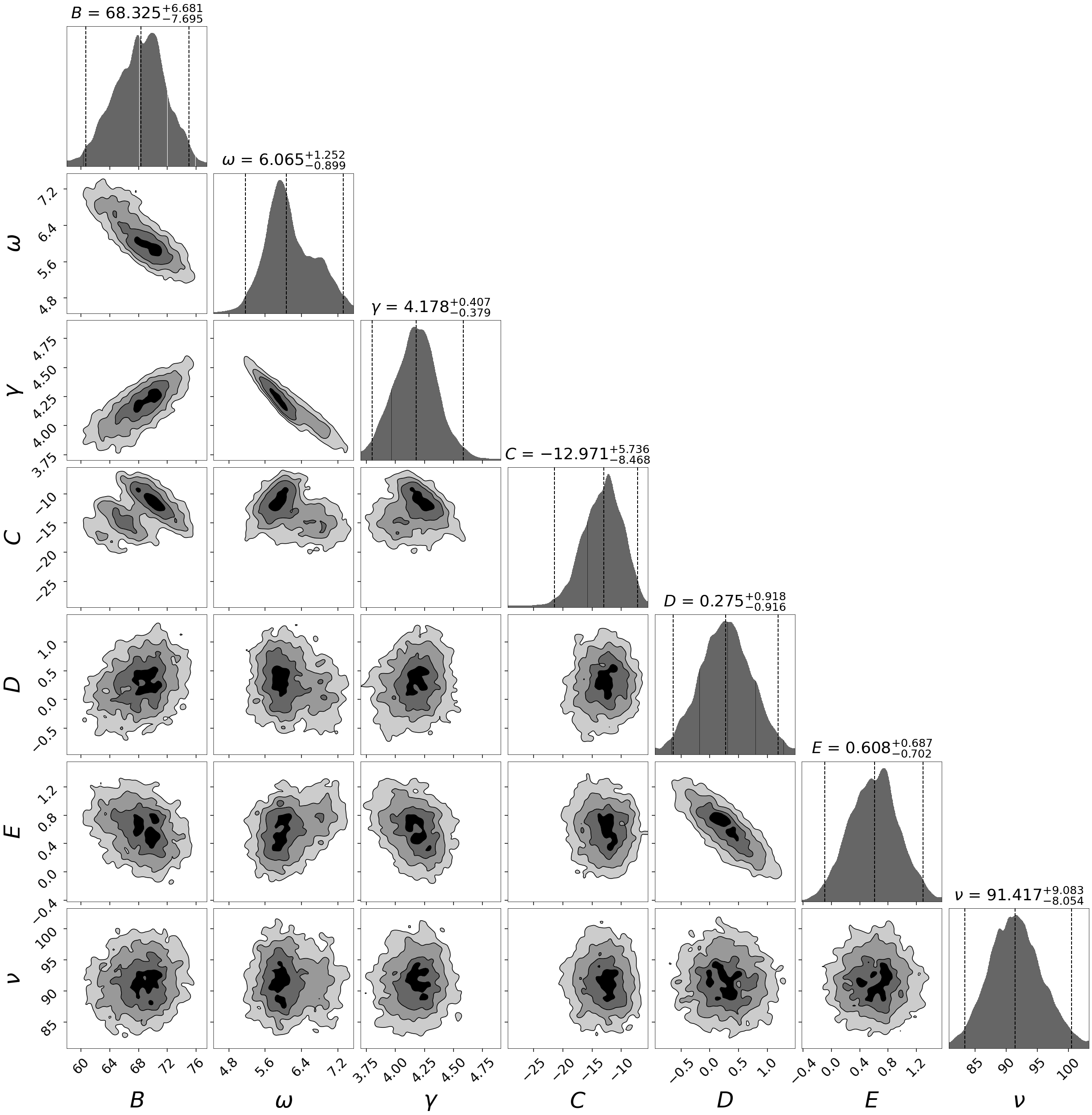} 
\caption{A cornerplot of the model parameters for the wavy midplane analysis using the epicyclic approximation and applied to the unclustered Radcliffe Wave young star sample. In the 1D distributions, the vertical dashed lines indicate the median and $2\sigma$ errors while in the 2D distributions, we demarcate the $0.5\sigma$, $1\sigma$, $1.5\sigma$, and $2\sigma$ contours. See Table \ref{tab:priors} for a description of the parameters and associated priors.}
\label{fig:cornerplot_epicyclic_unclustered}
\end{center}
\end{figure}

We fit this trend in vertical angle apparent in Figure \ref{fig:big3} using the model described in \S\ref{subsec:model} and the $\tt\string dynesty$ package. A summary of our results is shown in Table \ref{tab:bfparams}, which includes results for both the ``clustered" and ``unclustered" Radcliffe Wave samples that were fit using both the epicyclic and St\"ackel-Fudge approximations. We also include a corner plot for the unclustered young star sample using the epicyclic approximation, which is shown in Figure \ref{fig:cornerplot_epicyclic_unclustered}. Additional plots for the other three cases can be found in Appendix \ref{sec:cornerplots}. In Appendix \ref{sec:cornerplots}, we also show the results for the epicyclic fit to the main sequence control sample; the control sample fit is unconstrained, spanning a range of $3-220$ pc with most of the probability for the amplitude $B$ consistent with being $< 10$ pc, in contrast to the younger sample fits.

Our best-fit amplitude for the unclustered sample based on the epicyclic approximation is $B$ = 68 pc. However, we find a significantly higher amplitude of $B$ = 96 pc when only the clustered sample is fit, implying that potential interlopers in the unclustered sample have likely been removed. We find good agreement between the epicyclic and St\"ackel-Fudge best-fit parameters, with an estimated amplitude of $B$ = 65 pc and $B$ = 79 pc, respectively, for the unclustered and clustered St\"ackel-Fudge samples. 

We compute the best-fit period of the stellar wave using the best-fit angular frequency $\omega$ parameter, where $P = \frac{2\pi}{\omega}$. We find a best-fit period $P$ = 1.0 kpc for the epiycyclic unclustered sample (or $P$ = 1.1 kpc for the corresponding St\"ackel-Fudge approach). Just like the amplitude, the period also increases when restricting the fit to the clustered stars: we obtain a period of $P$ = 1.4 kpc using the epicyclic approximation, or $P$ = 1.2 kpc when using St\"ackel-Fudge approach.

Beyond the amplitude and period, we find a small value of the parameter $C$ (quantifying the offset of the midplane), which lies at $C\approx -12$ pc for three of the four fits. The $D$ parameter quantifies the deviation from the average velocity $E$ of the wave. Both of these values are also very small across all fits, on the order of $\lesssim \rm  1 \; km \; s^{-1}$, implying there is not a significant velocity gradient or velocity offset of the wavy midplane.

\begin{figure}[h!]
\begin{center}
\includegraphics[width=0.85\columnwidth]{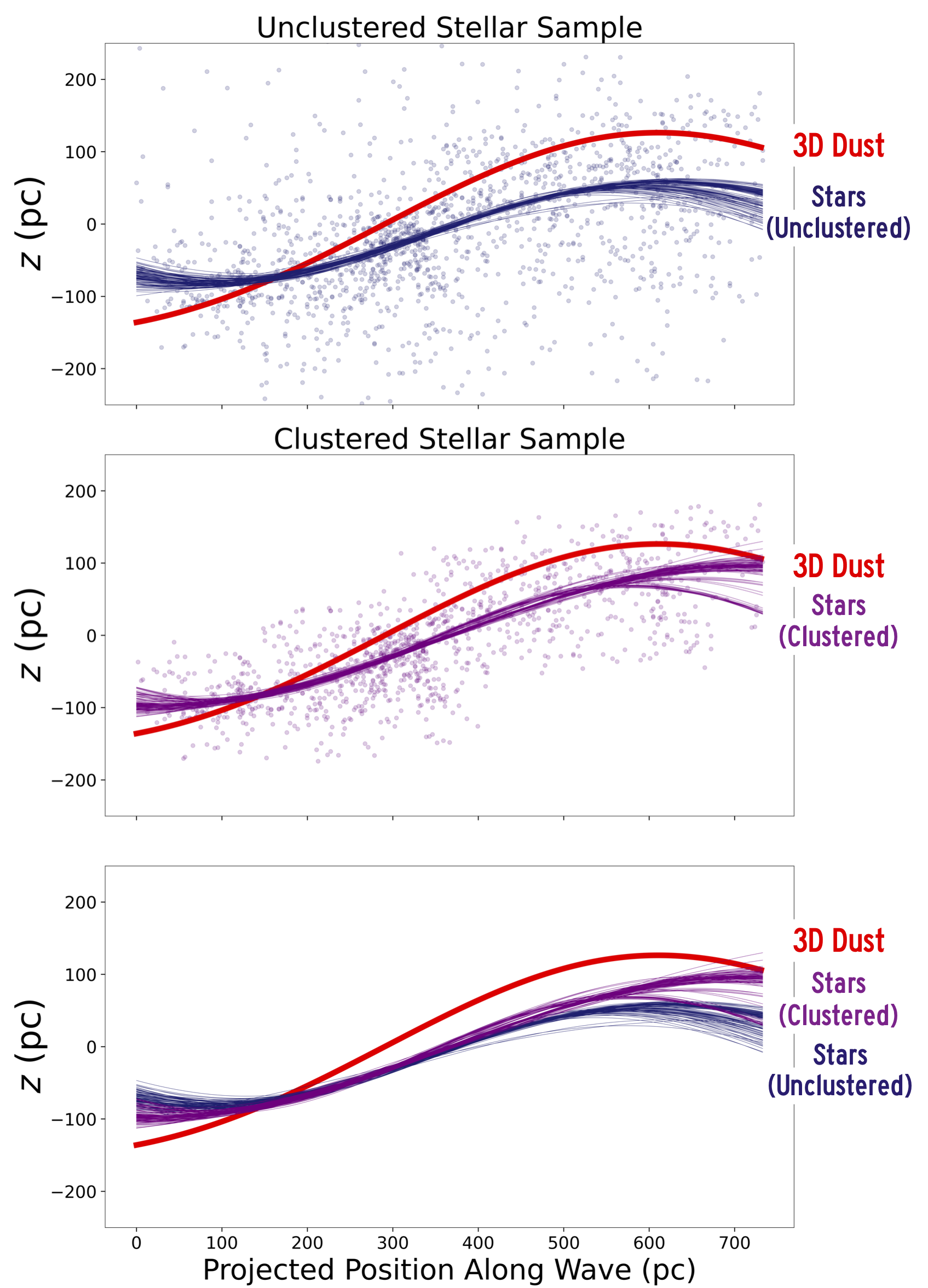}
\caption{Comparison of the Radcliffe Wave best-fit models derived using 3D dust \citet{Alves_2020} and this work. The top panel shows a comparison between the 3D dust Radcliffe Wave (red) and the epicyclic fit to the \textit{unclustered} stellar wave sample (navy). The middle panel shows a comparison between the 3D dust Radcliffe Wave (red) and the epicyclic fit to the \textit{clustered} stellar sample (indigo). In the bottom panel, we compare all three model fits.}
\label{fig:stellarvdust}
\end{center}
\end{figure}

\section{Discussion} \label{sec:discussion}

The ultimate goal of this work is two-fold. First, we seek to understand the differences between the Radcliffe Wave detected in 3D dust and young stars. And second, we seek to lay the foundation for understanding the Radcliffe Wave's origin via a detailed study of its 3D kinematics. 

\subsection{Comparison between Stellar and Gaseous Wave}

In Figure \ref{fig:stellarvdust} we show a comparison between the best-fit model for the Radcliffe Wave defined by 3D dust, as well as our best-fit model for the Radcliffe Wave defined using the epicylic fits to the unclustered and clustered Radcliffe Wave stellar samples. For the latter, we overlay 100 realizations of the stellar sinusoidal model, based on the best-fit parameters and their corresponding errors returned by {\tt\string dynesty}. 

We find that the amplitude and period of our best-fit model to the stellar wave are smaller than, but still consistent with, the  amplitude of the Radcliffe Wave detected in 3D dust. Specifically, in the 3D dust (fit to the distribution of molecular clouds), \citet{Alves_2020} obtain an average amplitude and period of 160 pc  and 1.9 kpc respectively when computed across the entire structure. However, computing the period and amplitude in the section of the wave overlapping with the \citet{Zari_2018} sample using the dampled sinusoidal model from \citet{Alves_2020}, we obtain a dust-based amplitude of $B$ = 140 pc and a period $P$ = 1.5 kpc in our region of interest. Compared to the best-fit amplitude and period of $B$ = 96 pc and $P$ = 1.4 kpc derived using the clustered epicyclic results, we find overall good agreement between the stellar and interstellar Radcliffe Wave. The smaller than expected amplitude in the stellar wave could be intrinsic, in that the gas is physically leading or trailing the stars in our region of interest.  However, the offset could also be due to the range of the Radcliffe Wave over which we fit: including stars only out to 500 pc in distance means our sample barely reaches the trough and crest of the wave compared to the 3D dust, which could bias our results toward lower amplitudes. 

Bolstering these results, wave-like structures with similar amplitudes and periods have been detected in the dust lanes of external galaxies. For example, \citet{Matthews_2008} and \citet{Narayan_2020} detect corrugations in several edge-on disk galaxies, which have amplitudes between $\approx 70 - 200$ pc and periods between $\approx 1-5$ kpc, consistent with both the stellar Radcliffe Wave constrained here and the gaseous Radcliffe Wave from \citet{Alves_2020}.

\subsection{Origin of the Radcliffe Wave}

The origin of the Radcliffe Wave could stem from a number of physical processes that can induce undulations in the disks of spiral galaxies. These processes can be external or internal to the disk. 

Potential external causes include perturbers (dwarf galaxies, dark matter subhaloes) and/or tidal interactions with companion galaxies \citep[e.g.][]{Lopez_2002,Widrow_2014}. For example, past studies have discovered wave-like asymmetries in the Galactic disk by analyzing both the number density of stars and their vertical velocities at varying $z$ \citep{Widrow_2012, Bennett_2019}, finding that a single external perturber event (occurring between 300 and 900 Myr) is responsible \citep{Antoja_2018} for the disk asymmetries in the solar neighborhood.  The favored hypothesis for such a perturbation is a fly-by from a satellite. Various simulations have reproduced the Milky Way's spiral arms and vertical oscillations with the Sagittarius dwarf galaxy (Sgr), under the constraint that the initial mass of Sgr is more massive than previously assumed \citep{Laporte_2019, Bland_Hawthorn_2019}. In contrast, \citet{Bennett_2021} have argued through a suite of N-body simulations that Sgr could not have produced the vertical waves in \citet{Bennett_2019}. 

Building on the work of both \citet{Bennett_2019} and \citet{Antoja_2018}, \citet{Thula_2021} explore the possibility that the Radcliffe Wave is indeed associated with a large-scale vertical kinematic oscillation in the disk. Based on a sample of open clusters, Upper Main Sequence (UMS) stars, OB stars, and Red Giant Branch (RGB) stars co-spatial with the Radcliffe Wave, \citet{Thula_2021} find a consistent kinematic wave-like behavior in the vertical velocity (on the order of a few $\rm km\;s\;^{-1}$) in the UMS, OB, and open cluster population, but not in the older RGB population. Running N-body simulations of a satellite as massive as the Sgr dwarf galaxy impacting the galactic disk, \citet{Thula_2021} conclude that the resultant density wave aligns with the Radcliffe Wave, suggesting the wave may be the result of an external perturber. However, \citet{Thula_2021} also note a misalignment between this density wave and the resulting kinematic wave, which will require further follow-up work with Gaia DR3. 

In comparison to \citet{Thula_2021}, we explore a much younger stellar population (ages of $<20$ Myr) as opposed to $\lesssim 100$ Myr in that work. We find no sinusoidal trend in vertical angle in the older main sequence population. In the case of an external perturber, it is possible that we are simply less sensitive to trends in vertical angle in the main sequence control sample, because the older stars have relaxed in comparison to the younger, dynamically colder pre-main sequence population. However, the trend in vertical angle in the youngest stellar populations $<20$ Myr most strongly associated with the gas could also favor more transitory and/or recent dynamical processes for the formation of the Radcliffe Wave. These transitory processes could include physical mechanisms internal to the disk, which, unlike a perturber, would only affect the motion of the gas, as opposed to all main sequence stars in the vicinity of the Radcliffe Wave. The pre-main sequence stars would then inherit the structure and motion of its natal material, accounting for this discrepancy in vertical angle between the younger and older stellar populations. 

For example, in the spirit of \citet{Nelson_1995}, \citet{Fleck_2020} suggests a Kelvin-Helmholtz fluid instability (KHI) arising from the cross-rotation of (i.e. difference in angular velocity between) gas in the Galactic disk and halo could give rise to a wave-like structure. Another scenario includes clustered supernova feedback, which simulations have shown are capable of pushing gas hundreds of parsecs out of the plane \citep{Kim_2018}. However, the supernova-driven scenario would require a high degree of coordination along the wave's 2.7 kpc length, rendering this mechanism somewhat less plausible than the KHI. 

\subsection{Velocity Variations along the Radcliffe Wave }
Independent of the origin for the spatial undulation in the Radcliffe Wave, we do not detect any bulk velocity structure of the gas, as evidenced by the small values of the $D$ (quantifying the velocity offset of the midplane) and $E$ (quantifying the deviation from the average velocity) parameters in Table \ref{tab:bfparams}. The fact that we do not detect any additional velocity structure suggests that the velocity variation must be on the order of a few $\rm km \; s^{-1}$, which is consistent with \citet{Thula_2021}. The small scale of the velocity variations is also consistent with the maximum vertical velocity of stars at the midplane of $\rm \lesssim 10 \; km\;s^{-1}$ predicted by the epicyclic approximation in Equation \ref{eq:vz} ($v_{z_{max}} \approx A_z \times \nu$) adopting the restoring frequency $\nu$ from Table \ref{tab:bfparams} and an amplitude of the stellar orbit, $A_Z$, equal to the amplitude of the stellar Radcliffe Wave. Any proposed origin for the Radcliffe Wave should thus not require vertical velocity variations across the wave of $\rm >10 \; km\;s^{-1}$.  Future observations will allow for more sensitive determinations of the velocity variation. \textit{Gaia} DR3 includes a factor of four more stars with radial velocities compared to our current \textit{Gaia} DR2 radial velocity sample, and these radial velocities are accompanied by smaller uncertainties. Likewise, recent work exploring the detailed structure of the Radcliffe Wave using complementary tracers, including OB stars \citep{Pantaleoni_2021} and the youngest open clusters \citep{Donada_2021}, provide additional means for understanding the kinematics of the Radcliffe Wave using different stellar populations. The combination of all of the above will enable us to revisit these results in more detail, mitigating the effect of the small sample size and large radial velocity errors in our current analysis.

\section{Conclusion} \label{sec:conclusion}

We leverage a sample of $\approx$ 1500 young, pre-main sequence stars \citep{Zari_2018} with radial velocity measurements from \textit{Gaia} to explore the spatial and kinematic structure of the Radcliffe Wave -- a 2.7 kpc band of molecular clouds recently discovered via 3D dust mapping \citet{Alves_2020}. Our main conclusions are as follows: 

\begin{itemize}

\item  We convert the stars' positions and velocities to action-angle coordinates and explore trends in the stars' vertical angle ($\Omega_z$) versus their position along the Radcliffe Wave. We uncover a significant trend in vertical angle indicative of a wave-like oscillation, which is not present in a control sample of older main sequence stars. The lack of a trend in vertical angle for older stars may favor a younger dynamical interpretation for the wave's origin.

\item We fit a sinusoidal model to the vertical angle data for the Radcliffe Wave's young stars. We compare the best-fit results for the stars with the known parameters of the gaseous Radcliffe Wave detected in 3D dust: we find good morphological agreement, albeit with the stellar data favoring a smaller amplitude and period than the 3D-dust-based results.

\item We do not detect significant variation in the vertical velocities as a function of position along the wave, suggesting that such variations would need to be smaller than a few kilometers per second. Future work incorporating the abundance of radial velocities in \textit{Gaia} DR3 will enable more sensitive constraints on the variation in the vertical velocities and its implications for the origin of the Radcliffe Wave.

\end{itemize}


\acknowledgments
The authors would like to thank the anonymous referee, whose thoughtful reading of our manuscript improved the quality of this work. AJT would like to acknowledge support from the Harvard PRISE Fellowship. CZ acknowledges that support for this work was provided by NASA through the NASA Hubble Fellowship grant HST-HF2-51498.001 awarded by the Space Telescope Science Institute, which is operated by the Association of Universities for Research in Astronomy, Inc., for NASA, under contract NAS5-26555. JSS remains grateful to Rebecca Bleich for her continued love and support. The visualization, exploration, and interpretation of data presented in this work were made possible using the glue visualization software, supported under NSF grant numbers OAC-1739657 and CDS\&E:AAG-1908419. AAG and CZ acknowledge support by NASA ADAP grant 80NSSC21K0634 “Knitting Together the Milky Way: An Integrated Model of the Galaxy’s Stars, Gas, and Dust.” AB acknowledges support by the Excellence Cluster ORIGINS which is funded by the German Research Foundation (DFG) under Germany’s Excellence Strategy EXC-2094-390783311.


\software{astropy \citep{2013A&A...558A..33A},  
          dynesty \citep{Speagle_2020},
          gala \citep{Price-Whelan_2017},
          pyia \citep{Price-Whelan_2018},
          galpy \citep{2015ApJS..216...29B},
          scikit \citep{scikit-learn},
          matplotlib \citep{Hunter:2007},
          glue \citep{glueviz}
          }



\appendix

\begin{figure}
\begin{center}
\includegraphics[width=0.75\columnwidth]{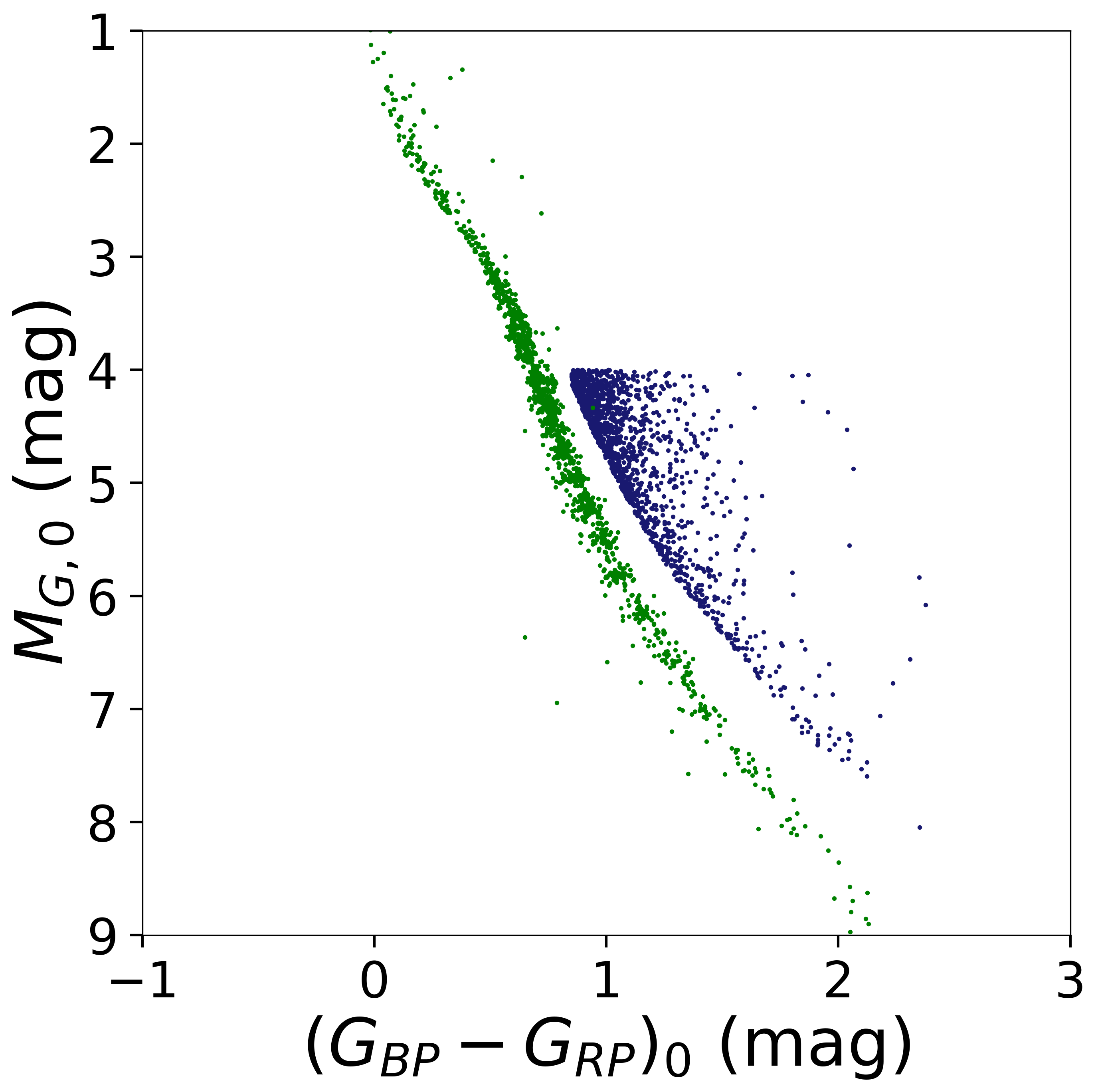}
\caption{Dereddened \textit{Gaia}-based color-magnitude diagram showing the primary unclustered pre-main sequence sample (navy) alongside the main sequence control sample (green). The sharp cutoff in $M_G,0$ for the pre-main sequence sample is stipulated in \citet{Zari_2018} and motivated by the need to exclude sources that are located on the main-sequence turn-off and on the faint end of the giant branch. \label{fig:cmd}}
\end{center}
\end{figure}

\section{Color-Magnitude Diagrams of Young and Old Stellar Samples} \label{sec:CMD}


To demonstrate that our control sample truly consists of older stars, we show a dereddened \textit{Gaia}-based color-magnitude diagram (CMD) in Figure \ref{fig:cmd}. As expected, there is a clear differentiation between the young (PMS) stars and the older control sample, with the PMS sample showing redder $BP-RP$ colors at fixed $G$ band absolute magnitude.

\section{Comparisons of action-angle approximation methods: {\tt\string gala} orbit integration vs. epicyclic approximation} \label{sec:epicomp}

The {\tt\string gala} package uses a computationally expensive algorithm to calculate actions and angles. However, a simpler, albeit less precise, method relies on the epicyclic approximation, which we use for fitting the sinusoidal model described in \S \ref{sec:methods}. Here we compare the two methods to determine whether the epicyclic approximation is performing well. Setting the offsets $z_0$ and $v_{z_0}$ to zero, Equation \ref{eq:omegaz} tells us that vertical angle is approximately:

\begin{equation}
\Omega_{z,\text{epi}} \approx \tan^{-1}\left(\frac{z}{v_z} \nu \right).
\end{equation}

\begin{figure}
\begin{center}
\includegraphics[width=1.0\columnwidth]{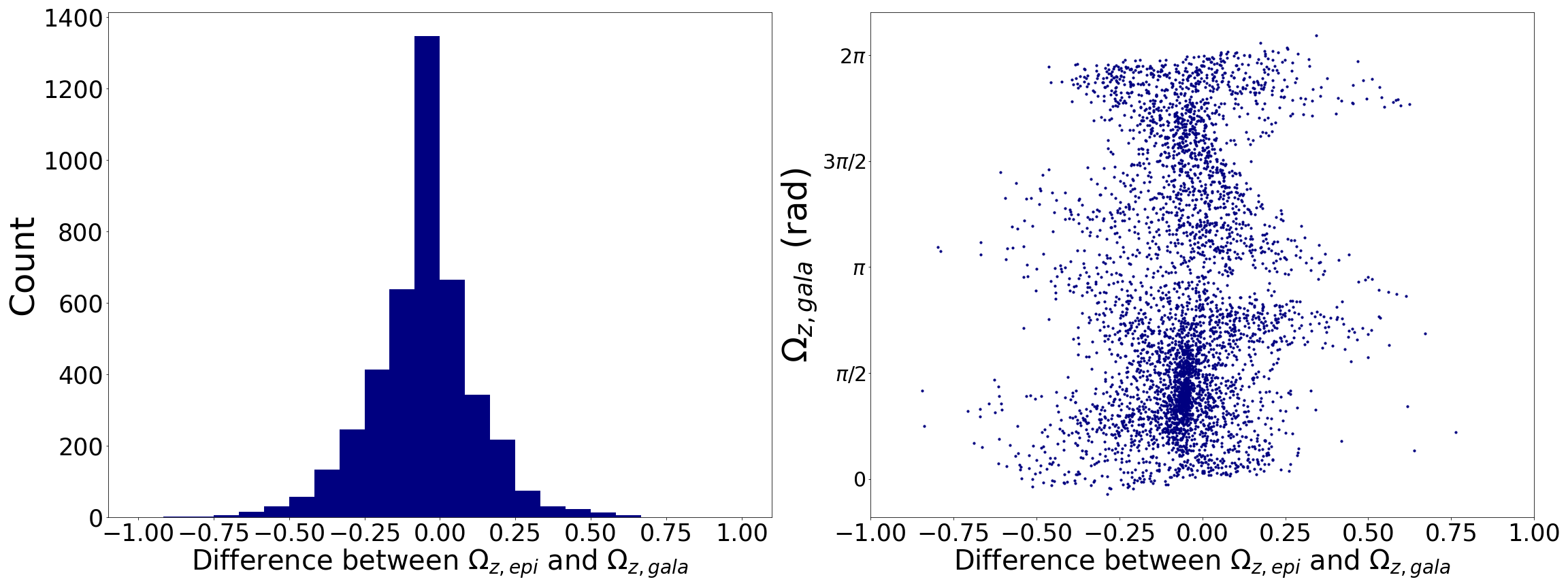} 
\caption{Left: Distribution of the difference in vertical angle value between the epicyclic approximation and direct integration of the orbit with {\tt\string gala} for each of the stars with radial velocities in the unclustered pre-main sequence sample from \citet{Zari_2018}. A difference of zero denotes a perfect match. Right: The difference instead plotted against the vertical angle calculated from \texttt{gala}, highlighting that the greatest discrepancies occur near $0$, $\pi$, and $2\pi$.}
\label{fig:gala}
\end{center}
\end{figure}

The \texttt{galpy} package gives a value of $\nu \approx 75 \; \rm km \; s^{-1}$\;kpc$^{-1}$ as the default vertical frequency of its {\tt\string MWPotential2014}. For each star in the Radcliffe sample, we compute the epicyclic value $\Omega_{\rm z,epi}$, then compare it to the {\tt\string gala}-computed value $\Omega_{z,\text{gala}}$. The ratios are shown in a histogram in Figure \ref{fig:gala}. We find a that the histogram peaks around zero with a relatively small scatter ($\lesssim 0.25$) indicating that the epicyclic approach is consistent with the more expensive \texttt{gala} approach for our sample. 

\section{Additional Wavy Midplane Model Results} \label{sec:cornerplots}

\begin{figure}[ht!]
\begin{center}
\includegraphics[width=1.0\columnwidth]{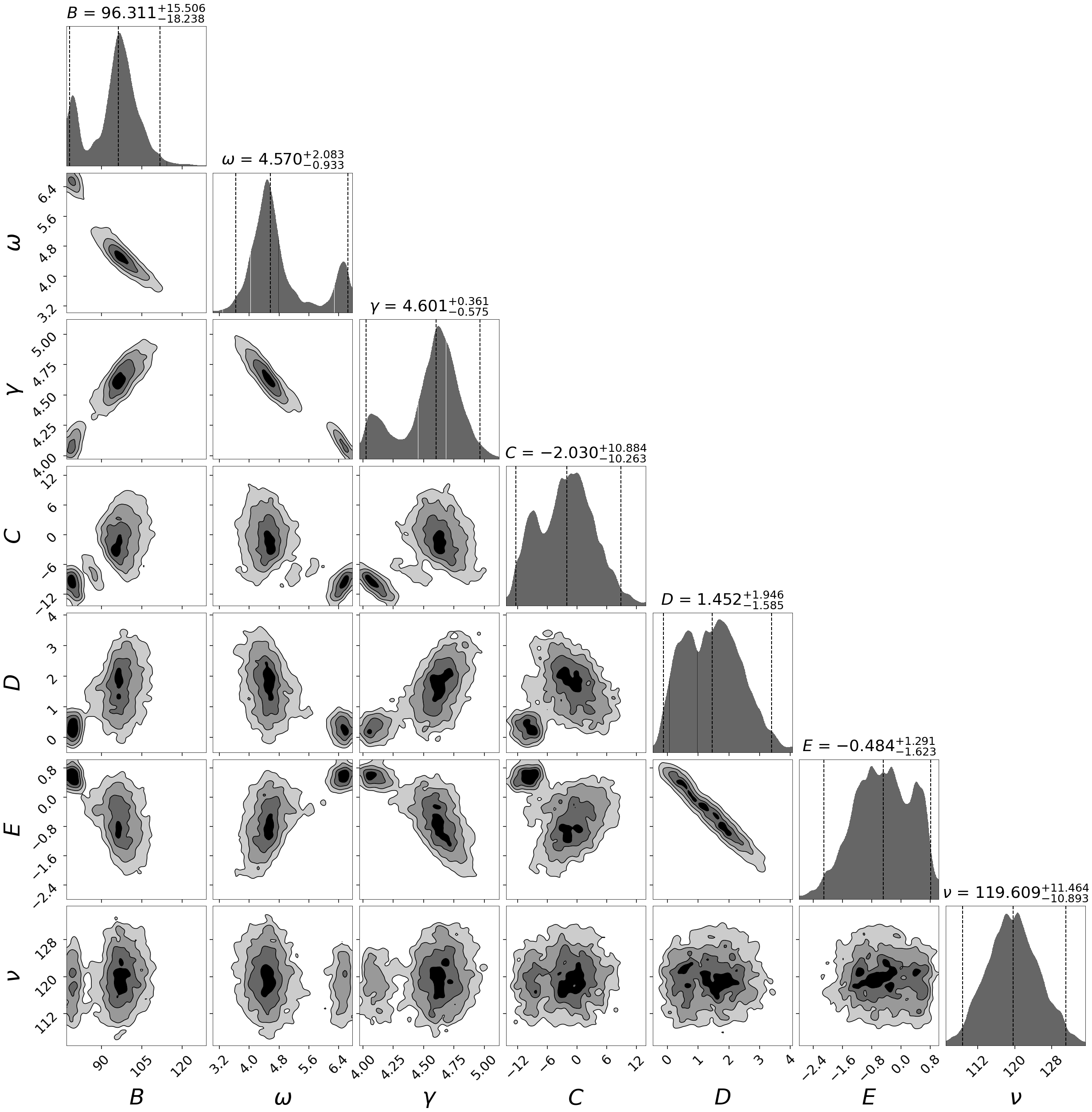} 
\caption{Same as Figure \ref{fig:cornerplot_epicyclic_unclustered}, but for the epicyclic approximation applied to the clustered Radcliffe Wave sample.}
\label{fig:cornerplot_epicyclic_clustered}
\end{center}
\end{figure}

\begin{figure}[ht!]
\begin{center}
\includegraphics[width=1.0\columnwidth]{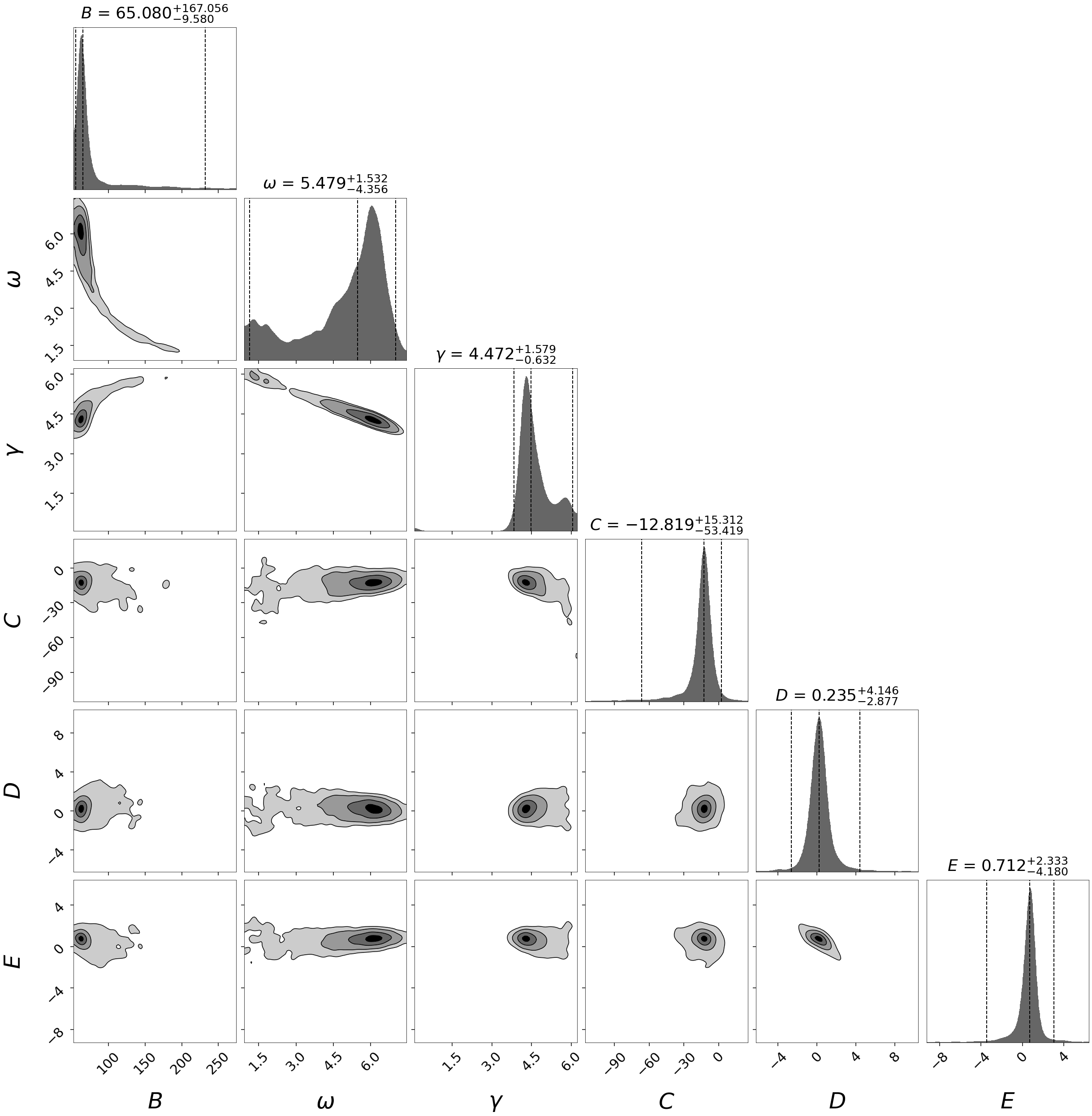} 
\caption{Same as Figure \ref{fig:cornerplot_epicyclic_unclustered}, but for the St\"ackel-Fudge approach applied to the unclustered Radcliffe Wave sample.}
\label{fig:cornerplot_staeckel_unclustered}
\end{center}
\end{figure}

\begin{figure}[ht!]
\begin{center}
\includegraphics[width=1.0\columnwidth]{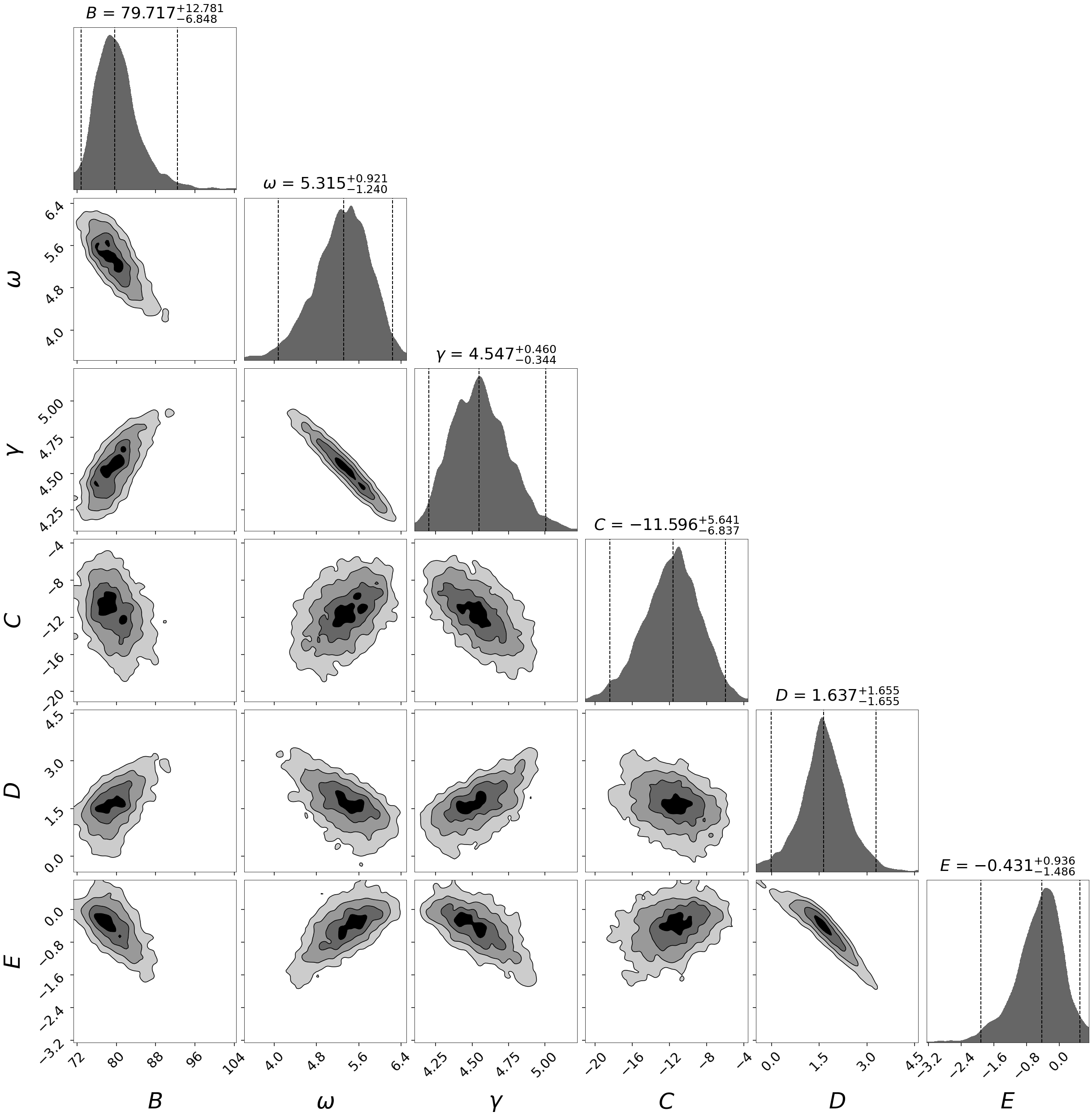} 
\caption{Same as Figure \ref{fig:cornerplot_epicyclic_unclustered}, but for the St\"ackel-Fudge approach applied to the clustered Radcliffe Wave sample.}
\label{fig:cornerplot_staeckel_clustered}
\end{center}
\end{figure}

\begin{figure}[ht!]
\begin{center}
\includegraphics[width=1.0\columnwidth]{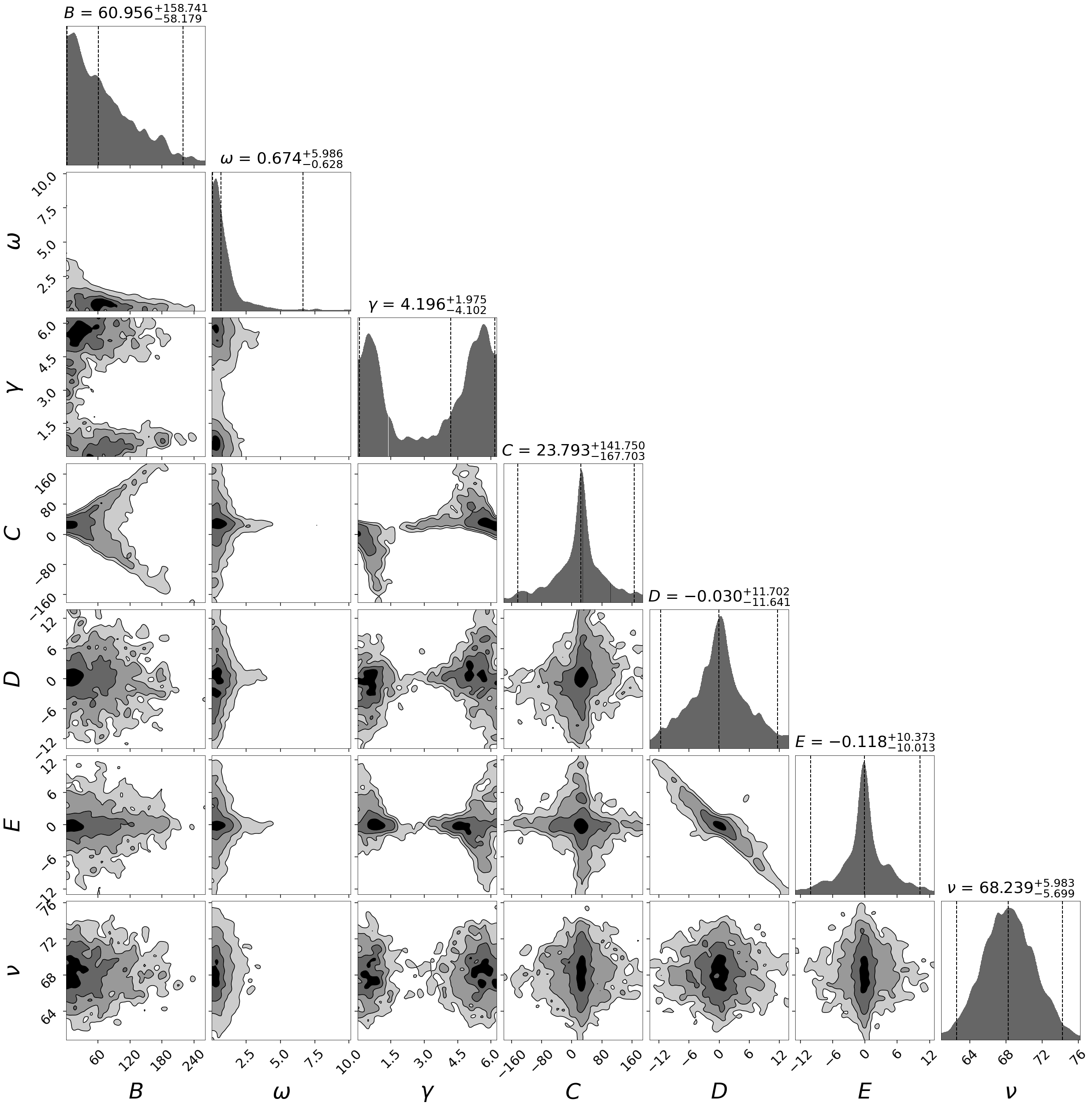} 
\caption{Same as Figure \ref{fig:cornerplot_epicyclic_unclustered}, but for the epicyclic approach applied to a random control sample of older, main sequence stars occupying the same volume as the younger, Radcliffe Wave sample.}
\label{fig:cornerplot_epicyclic_control}
\end{center}
\end{figure}

In the main text we summarized the overall parameter estimates derived from our four model fits to the young, pre-main sequence Radcliffe Wave sample and highlighted a summary of the results using the epicyclic approximation over the unclustered sample in Figure \ref{fig:cornerplot_epicyclic_unclustered}. The corner plots for the epicyclic approximation over the clustered sample, the St\"ackel-Fudge approximation over the unclustered sample, and the St\"ackel-Fudge approximation over the clustered sample can be found in Figures \ref{fig:cornerplot_epicyclic_clustered}, \ref{fig:cornerplot_staeckel_unclustered}, and Figure \ref{fig:cornerplot_staeckel_clustered}, respectively. In Figure \ref{fig:cornerplot_epicyclic_control}, we also show the results of fitting the wavy midplane model to a sample of randomly selected, older main sequence stars occupying the same volume as the younger Radcliffe Wave sample. Unlike for the younger Radcliffe Wave sample, we find the control sample fit to be unconstrained and encompassing the most of the prior bounds. For example, $2\sigma$ error range on the amplitude $B$ spans $3 - 220$ pc, with most of the probability consistent with an amplitude $B < 10$ pc. 

\bibliography{sample63}{}

\begin{thebibliography}{}
\expandafter\ifx\csname natexlab\endcsname\relax\def\natexlab#1{#1}\fi
\providecommand{\url}[1]{\href{#1}{#1}}
\providecommand{\dodoi}[1]{doi:~\href{http://doi.org/#1}{\nolinkurl{#1}}}
\providecommand{\doeprint}[1]{\href{http://ascl.net/#1}{\nolinkurl{http://ascl.net/#1}}}
\providecommand{\doarXiv}[1]{\href{https://arxiv.org/abs/#1}{\nolinkurl{https://arxiv.org/abs/#1}}}

\bibitem[{{Alves} {et~al.}(2020){Alves}, {Zucker}, {Goodman}, {Speagle},
  {Meingast}, {Robitaille}, {Finkbeiner}, {Schlafly}, \& {Green}}]{Alves_2020}
{Alves}, J., {Zucker}, C., {Goodman}, A.~A., {et~al.} 2020, \nat, 578, 237,
  \dodoi{10.1038/s41586-019-1874-z}

\bibitem[{{Antoja} {et~al.}(2018){Antoja}, {Helmi}, {Romero-G{\'o}mez}, {Katz},
  {Babusiaux}, {Drimmel}, {Evans}, {Figueras}, {Poggio}, {Reyl{\'e}}, {Robin},
  {Seabroke}, \& {Soubiran}}]{Antoja_2018}
{Antoja}, T., {Helmi}, A., {Romero-G{\'o}mez}, M., {et~al.} 2018, \nat, 561,
  360, \dodoi{10.1038/s41586-018-0510-7}

\bibitem[{{Astropy Collaboration} {et~al.}(2013){Astropy Collaboration},
  {Robitaille}, {Tollerud}, {Greenfield}, {Droettboom}, {Bray}, {Aldcroft},
  {Davis}, {Ginsburg}, {Price-Whelan}, {Kerzendorf}, {Conley}, {Crighton},
  {Barbary}, {Muna}, {Ferguson}, {Grollier}, {Parikh}, {Nair}, {Unther},
  {Deil}, {Woillez}, {Conseil}, {Kramer}, {Turner}, {Singer}, {Fox}, {Weaver},
  {Zabalza}, {Edwards}, {Azalee Bostroem}, {Burke}, {Casey}, {Crawford},
  {Dencheva}, {Ely}, {Jenness}, {Labrie}, {Lim}, {Pierfederici}, {Pontzen},
  {Ptak}, {Refsdal}, {Servillat}, \& {Streicher}}]{2013A&A...558A..33A}
{Astropy Collaboration}, {Robitaille}, T.~P., {Tollerud}, E.~J., {et~al.} 2013,
  \aap, 558, A33, \dodoi{10.1051/0004-6361/201322068}

\bibitem[{{Astropy Collaboration} {et~al.}(2018){Astropy Collaboration},
  {Price-Whelan}, {Sip{\H{o}}cz}, {G{\"u}nther}, {Lim}, {Crawford}, {Conseil},
  {Shupe}, {Craig}, {Dencheva}, {Ginsburg}, {VanderPlas}, {Bradley},
  {P{\'e}rez-Su{\'a}rez}, {de Val-Borro}, {Aldcroft}, {Cruz}, {Robitaille},
  {Tollerud}, {Ardelean}, {Babej}, {Bach}, {Bachetti}, {Bakanov}, {Bamford},
  {Barentsen}, {Barmby}, {Baumbach}, {Berry}, {Biscani}, {Boquien}, {Bostroem},
  {Bouma}, {Brammer}, {Bray}, {Breytenbach}, {Buddelmeijer}, {Burke},
  {Calderone}, {Cano Rodr{\'\i}guez}, {Cara}, {Cardoso}, {Cheedella}, {Copin},
  {Corrales}, {Crichton}, {D'Avella}, {Deil}, {Depagne}, {Dietrich}, {Donath},
  {Droettboom}, {Earl}, {Erben}, {Fabbro}, {Ferreira}, {Finethy}, {Fox},
  {Garrison}, {Gibbons}, {Goldstein}, {Gommers}, {Greco}, {Greenfield},
  {Groener}, {Grollier}, {Hagen}, {Hirst}, {Homeier}, {Horton}, {Hosseinzadeh},
  {Hu}, {Hunkeler}, {Ivezi{\'c}}, {Jain}, {Jenness}, {Kanarek}, {Kendrew},
  {Kern}, {Kerzendorf}, {Khvalko}, {King}, {Kirkby}, {Kulkarni}, {Kumar},
  {Lee}, {Lenz}, {Littlefair}, {Ma}, {Macleod}, {Mastropietro}, {McCully},
  {Montagnac}, {Morris}, {Mueller}, {Mumford}, {Muna}, {Murphy}, {Nelson},
  {Nguyen}, {Ninan}, {N{\"o}the}, {Ogaz}, {Oh}, {Parejko}, {Parley}, {Pascual},
  {Patil}, {Patil}, {Plunkett}, {Prochaska}, {Rastogi}, {Reddy Janga},
  {Sabater}, {Sakurikar}, {Seifert}, {Sherbert}, {Sherwood-Taylor}, {Shih},
  {Sick}, {Silbiger}, {Singanamalla}, {Singer}, {Sladen}, {Sooley},
  {Sornarajah}, {Streicher}, {Teuben}, {Thomas}, {Tremblay}, {Turner},
  {Terr{\'o}n}, {van Kerkwijk}, {de la Vega}, {Watkins}, {Weaver}, {Whitmore},
  {Woillez}, {Zabalza}, \& {Astropy Contributors}}]{astropy}
{Astropy Collaboration}, {Price-Whelan}, A.~M., {Sip{\H{o}}cz}, B.~M., {et~al.}
  2018, \aj, 156, 123, \dodoi{10.3847/1538-3881/aabc4f}

\bibitem[{{Beane} {et~al.}(2018){Beane}, {Ness}, \& {Bedell}}]{Beane_2018}
{Beane}, A., {Ness}, M.~K., \& {Bedell}, M. 2018, \apj, 867, 31,
  \dodoi{10.3847/1538-4357/aae07f}

\bibitem[{{Beane} {et~al.}(2019){Beane}, {Sanderson}, {Ness}, {Johnston},
  {Grion Filho}, {Mac Low}, {Angl{\'e}s-Alc{\'a}zar}, {Hogg}, \&
  {Laporte}}]{Beane_2019}
{Beane}, A., {Sanderson}, R.~E., {Ness}, M.~K., {et~al.} 2019, \apj, 883, 103,
  \dodoi{10.3847/1538-4357/ab3d3c}

\bibitem[{{Beaumont} {et~al.}(2015){Beaumont}, {Goodman}, \&
  {Greenfield}}]{glueviz}
{Beaumont}, C., {Goodman}, A., \& {Greenfield}, P. 2015, in Astronomical
  Society of the Pacific Conference Series, Vol. 495, Astronomical Data
  Analysis Software an Systems XXIV (ADASS XXIV), ed. A.~R. {Taylor} \&
  E.~{Rosolowsky}, 101

\bibitem[{{Bennett} \& {Bovy}(2019)}]{Bennett_2019}
{Bennett}, M., \& {Bovy}, J. 2019, \mnras, 482, 1417,
  \dodoi{10.1093/mnras/sty2813}

\bibitem[{{Bennett} \& {Bovy}(2021)}]{Bennett_2021}
---. 2021, \mnras, 503, 376, \dodoi{10.1093/mnras/stab524}

\bibitem[{{Binney}(2012)}]{2012MNRAS.426.1324B}
{Binney}, J. 2012, \mnras, 426, 1324, \dodoi{10.1111/j.1365-2966.2012.21757.x}

\bibitem[{{Binney} \& {Tremaine}(2008)}]{Binney_2008}
{Binney}, J., \& {Tremaine}, S. 2008, {Galactic Dynamics: Second Edition}

\bibitem[{{Bland-Hawthorn} \& {Tepper-Garc{\'\i}a}(2021)}]{Bland_Hawthorn_2021}
{Bland-Hawthorn}, J., \& {Tepper-Garc{\'\i}a}, T. 2021, \mnras,
  \dodoi{10.1093/mnras/stab704}

\bibitem[{{Bland-Hawthorn} {et~al.}(2019){Bland-Hawthorn}, {Sharma},
  {Tepper-Garcia}, {Binney}, {Freeman}, {Hayden}, {Kos}, {De Silva}, {Ellis},
  {Lewis}, {Asplund}, {Buder}, {Casey}, {D'Orazi}, {Duong}, {Khanna}, {Lin},
  {Lind}, {Martell}, {Ness}, {Simpson}, {Zucker}, {Zwitter}, {Kafle},
  {Quillen}, {Ting}, \& {Wyse}}]{Bland_Hawthorn_2019}
{Bland-Hawthorn}, J., {Sharma}, S., {Tepper-Garcia}, T., {et~al.} 2019, \mnras,
  486, 1167, \dodoi{10.1093/mnras/stz217}

\bibitem[{{Bovy}(2015)}]{2015ApJS..216...29B}
{Bovy}, J. 2015, \apjs, 216, 29, \dodoi{10.1088/0067-0049/216/2/29}

\bibitem[{{Chen} {et~al.}(2020){Chen}, {D'Onghia}, {Alves}, \&
  {Adamo}}]{Chen_2020}
{Chen}, B., {D'Onghia}, E., {Alves}, J., \& {Adamo}, A. 2020, \aap, 643, A114,
  \dodoi{10.1051/0004-6361/201935955}

\bibitem[{{de Zeeuw}(1985)}]{1985MNRAS.216..273D}
{de Zeeuw}, T. 1985, \mnras, 216, 273, \dodoi{10.1093/mnras/216.2.273}

\bibitem[{{Deason} {et~al.}(2017){Deason}, {Belokurov}, {Erkal}, {Koposov}, \&
  {Mackey}}]{Deason_2017}
{Deason}, A.~J., {Belokurov}, V., {Erkal}, D., {Koposov}, S.~E., \& {Mackey},
  D. 2017, \mnras, 467, 2636, \dodoi{10.1093/mnras/stx263}

\bibitem[{{Donada} \& {Figueras}(2021)}]{Donada_2021}
{Donada}, J., \& {Figueras}, F. 2021, arXiv e-prints, arXiv:2111.04685.
\newblock \doarXiv{2111.04685}

\bibitem[{Dormand \& Prince(1980)}]{DORMAND198019}
Dormand, J., \& Prince, P. 1980, Journal of Computational and Applied
  Mathematics, 6, 19, \dodoi{https://doi.org/10.1016/0771-050X(80)90013-3}

\bibitem[{Ester {et~al.}(1996)Ester, Kriegel, Sander, \& Xu}]{Ester_1996}
Ester, M., Kriegel, H.-P., Sander, J., \& Xu, X. 1996, in Proceedings of the
  Second International Conference on Knowledge Discovery and Data Mining,
  KDD'96 (AAAI Press), 226–231

\bibitem[{{Fleck}(2020)}]{Fleck_2020}
{Fleck}, R. 2020, \nat, 583, E24, \dodoi{10.1038/s41586-020-2476-5}

\bibitem[{{Foster} {et~al.}(2015){Foster}, {Cottaar}, {Covey}, {Arce}, {Meyer},
  {Nidever}, {Stassun}, {Tan}, {Chojnowski}, {da Rio}, {Flaherty}, {Rebull},
  {Frinchaboy}, {Majewski}, {Skrutskie}, {Wilson}, \& {Zasowski}}]{Foster_2015}
{Foster}, J.~B., {Cottaar}, M., {Covey}, K.~R., {et~al.} 2015, \apj, 799, 136,
  \dodoi{10.1088/0004-637X/799/2/136}

\bibitem[{{Gaia Collaboration} {et~al.}(2018){Gaia Collaboration}, {Brown},
  {Vallenari}, {Prusti}, {de Bruijne}, {Babusiaux}, {Bailer-Jones}, {Biermann},
  {Evans}, {Eyer}, {Jansen}, {Jordi}, {Klioner}, {Lammers}, {Lindegren},
  {Luri}, {Mignard}, {Panem}, {Pourbaix}, {Randich}, {Sartoretti}, {Siddiqui},
  {Soubiran}, {van Leeuwen}, {Walton}, {Arenou}, {Bastian}, {Cropper},
  {Drimmel}, {Katz}, {Lattanzi}, {Bakker}, {Cacciari}, {Casta{\~n}eda},
  {Chaoul}, {Cheek}, {De Angeli}, {Fabricius}, {Guerra}, {Holl}, {Masana},
  {Messineo}, {Mowlavi}, {Nienartowicz}, {Panuzzo}, {Portell}, {Riello},
  {Seabroke}, {Tanga}, {Th{\'e}venin}, {Gracia-Abril}, {Comoretto},
  {Garcia-Reinaldos}, {Teyssier}, {Altmann}, {Andrae}, {Audard},
  {Bellas-Velidis}, {Benson}, {Berthier}, {Blomme}, {Burgess}, {Busso},
  {Carry}, {Cellino}, {Clementini}, {Clotet}, {Creevey}, {Davidson}, {De
  Ridder}, {Delchambre}, {Dell'Oro}, {Ducourant},
  {Fern{\'a}ndez-Hern{\'a}ndez}, {Fouesneau}, {Fr{\'e}mat}, {Galluccio},
  {Garc{\'\i}a-Torres}, {Gonz{\'a}lez-N{\'u}{\~n}ez}, {Gonz{\'a}lez-Vidal},
  {Gosset}, {Guy}, {Halbwachs}, {Hambly}, {Harrison}, {Hern{\'a}ndez},
  {Hestroffer}, {Hodgkin}, {Hutton}, {Jasniewicz}, {Jean-Antoine-Piccolo},
  {Jordan}, {Korn}, {Krone-Martins}, {Lanzafame}, {Lebzelter}, {L{\"o}ffler},
  {Manteiga}, {Marrese}, {Mart{\'\i}n-Fleitas}, {Moitinho}, {Mora}, {Muinonen},
  {Osinde}, {Pancino}, {Pauwels}, {Petit}, {Recio-Blanco}, {Richards},
  {Rimoldini}, {Robin}, {Sarro}, {Siopis}, {Smith}, {Sozzetti}, {S{\"u}veges},
  {Torra}, {van Reeven}, {Abbas}, {Abreu Aramburu}, {Accart}, {Aerts},
  {Altavilla}, {{\'A}lvarez}, {Alvarez}, {Alves}, {Anderson}, {Andrei},
  {Anglada Varela}, {Antiche}, {Antoja}, {Arcay}, {Astraatmadja}, {Bach},
  {Baker}, {Balaguer-N{\'u}{\~n}ez}, {Balm}, {Barache}, {Barata}, {Barbato},
  {Barblan}, {Barklem}, {Barrado}, {Barros}, {Barstow}, {Bartholom{\'e}
  Mu{\~n}oz}, {Bassilana}, {Becciani}, {Bellazzini}, {Berihuete}, {Bertone},
  {Bianchi}, {Bienaym{\'e}}, {Blanco-Cuaresma}, {Boch}, {Boeche}, {Bombrun},
  {Borrachero}, {Bossini}, {Bouquillon}, {Bourda}, {Bragaglia}, {Bramante},
  {Breddels}, {Bressan}, {Brouillet}, {Br{\"u}semeister}, {Brugaletta},
  {Bucciarelli}, {Burlacu}, {Busonero}, {Butkevich}, {Buzzi}, {Caffau},
  {Cancelliere}, {Cannizzaro}, {Cantat-Gaudin}, {Carballo}, {Carlucci},
  {Carrasco}, {Casamiquela}, {Castellani}, {Castro-Ginard}, {Charlot},
  {Chemin}, {Chiavassa}, {Cocozza}, {Costigan}, {Cowell}, {Crifo}, {Crosta},
  {Crowley}, {Cuypers}, {Dafonte}, {Damerdji}, {Dapergolas}, {David}, {David},
  {de Laverny}, {De Luise}, {De March}, {de Martino}, {de Souza}, {de Torres},
  {Debosscher}, {del Pozo}, {Delbo}, {Delgado}, {Delgado}, {Di Matteo},
  {Diakite}, {Diener}, {Distefano}, {Dolding}, {Drazinos}, {Dur{\'a}n},
  {Edvardsson}, {Enke}, {Eriksson}, {Esquej}, {Eynard Bontemps}, {Fabre},
  {Fabrizio}, {Faigler}, {Falc{\~a}o}, {Farr{\`a}s Casas}, {Federici},
  {Fedorets}, {Fernique}, {Figueras}, {Filippi}, {Findeisen}, {Fonti},
  {Fraile}, {Fraser}, {Fr{\'e}zouls}, {Gai}, {Galleti}, {Garabato},
  {Garc{\'\i}a-Sedano}, {Garofalo}, {Garralda}, {Gavel}, {Gavras}, {Gerssen},
  {Geyer}, {Giacobbe}, {Gilmore}, {Girona}, {Giuffrida}, {Glass}, {Gomes},
  {Granvik}, {Gueguen}, {Guerrier}, {Guiraud}, {Guti{\'e}rrez-S{\'a}nchez},
  {Haigron}, {Hatzidimitriou}, {Hauser}, {Haywood}, {Heiter}, {Helmi}, {Heu},
  {Hilger}, {Hobbs}, {Hofmann}, {Holland}, {Huckle}, {Hypki}, {Icardi},
  {Jan{\ss}en}, {Jevardat de Fombelle}, {Jonker}, {Juh{\'a}sz}, {Julbe},
  {Karampelas}, {Kewley}, {Klar}, {Kochoska}, {Kohley}, {Kolenberg},
  {Kontizas}, {Kontizas}, {Koposov}, {Kordopatis}, {Kostrzewa-Rutkowska},
  {Koubsky}, {Lambert}, {Lanza}, {Lasne}, {Lavigne}, {Le Fustec}, {Le
  Poncin-Lafitte}, {Lebreton}, {Leccia}, {Leclerc}, {Lecoeur-Taibi},
  {Lenhardt}, {Leroux}, {Liao}, {Licata}, {Lindstr{\o}m}, {Lister}, {Livanou},
  {Lobel}, {L{\'o}pez}, {Managau}, {Mann}, {Mantelet}, {Marchal}, {Marchant},
  {Marconi}, {Marinoni}, {Marschalk{\'o}}, {Marshall}, {Martino}, {Marton},
  {Mary}, {Massari}, {Matijevi{\v{c}}}, {Mazeh}, {McMillan}, {Messina},
  {Michalik}, {Millar}, {Molina}, {Molinaro}, {Moln{\'a}r}, {Montegriffo},
  {Mor}, {Morbidelli}, {Morel}, {Morris}, {Mulone}, {Muraveva}, {Musella},
  {Nelemans}, {Nicastro}, {Noval}, {O'Mullane}, {Ord{\'e}novic},
  {Ord{\'o}{\~n}ez-Blanco}, {Osborne}, {Pagani}, {Pagano}, {Pailler},
  {Palacin}, {Palaversa}, {Panahi}, {Pawlak}, {Piersimoni}, {Pineau}, {Plachy},
  {Plum}, {Poggio}, {Poujoulet}, {Pr{\v{s}}a}, {Pulone}, {Racero}, {Ragaini},
  {Rambaux}, {Ramos-Lerate}, {Regibo}, {Reyl{\'e}}, {Riclet}, {Ripepi}, {Riva},
  {Rivard}, {Rixon}, {Roegiers}, {Roelens}, {Romero-G{\'o}mez}, {Rowell},
  {Royer}, {Ruiz-Dern}, {Sadowski}, {Sagrist{\`a} Sell{\'e}s}, {Sahlmann},
  {Salgado}, {Salguero}, {Sanna}, {Santana-Ros}, {Sarasso}, {Savietto},
  {Schultheis}, {Sciacca}, {Segol}, {Segovia}, {S{\'e}gransan}, {Shih},
  {Siltala}, {Silva}, {Smart}, {Smith}, {Solano}, {Solitro}, {Sordo}, {Soria
  Nieto}, {Souchay}, {Spagna}, {Spoto}, {Stampa}, {Steele},
  {Steidelm{\"u}ller}, {Stephenson}, {Stoev}, {Suess}, {Surdej}, {Szabados},
  {Szegedi-Elek}, {Tapiador}, {Taris}, {Tauran}, {Taylor}, {Teixeira},
  {Terrett}, {Teyssand ier}, {Thuillot}, {Titarenko}, {Torra Clotet}, {Turon},
  {Ulla}, {Utrilla}, {Uzzi}, {Vaillant}, {Valentini}, {Valette}, {van Elteren},
  {Van Hemelryck}, {van Leeuwen}, {Vaschetto}, {Vecchiato}, {Veljanoski},
  {Viala}, {Vicente}, {Vogt}, {von Essen}, {Voss}, {Votruba}, {Voutsinas},
  {Walmsley}, {Weiler}, {Wertz}, {Wevers}, {Wyrzykowski}, {Yoldas},
  {{\v{Z}}erjal}, {Ziaeepour}, {Zorec}, {Zschocke}, {Zucker}, {Zurbach}, \&
  {Zwitter}}]{Gaia_2018}
{Gaia Collaboration}, {Brown}, A.~G.~A., {Vallenari}, A., {et~al.} 2018, \aap,
  616, A1, \dodoi{10.1051/0004-6361/201833051}

\bibitem[{{Galli} {et~al.}(2019){Galli}, {Loinard}, {Bouy}, {Sarro},
  {Ortiz-Le{\'o}n}, {Dzib}, {Olivares}, {Heyer}, {Hernandez},
  {Rom{\'a}n-Z{\'u}{\~n}iga}, {Kounkel}, \& {Covey}}]{Galli_2019}
{Galli}, P.~A.~B., {Loinard}, L., {Bouy}, H., {et~al.} 2019, \aap, 630, A137,
  \dodoi{10.1051/0004-6361/201935928}

\bibitem[{{Green} {et~al.}(2019){Green}, {Schlafly}, {Zucker}, {Speagle}, \&
  {Finkbeiner}}]{Green_2019}
{Green}, G.~M., {Schlafly}, E., {Zucker}, C., {Speagle}, J.~S., \&
  {Finkbeiner}, D. 2019, \apj, 887, 93, \dodoi{10.3847/1538-4357/ab5362}

\bibitem[{{Gro{\ss}schedl} {et~al.}(2021){Gro{\ss}schedl}, {Alves}, {Meingast},
  \& {Herbst-Kiss}}]{Grossschedl_2021}
{Gro{\ss}schedl}, J.~E., {Alves}, J., {Meingast}, S., \& {Herbst-Kiss}, G.
  2021, \aap, 647, A91, \dodoi{10.1051/0004-6361/202038913}

\bibitem[{{Hacar} {et~al.}(2016){Hacar}, {Alves}, {Forbrich}, {Meingast},
  {Kubiak}, \& {Gro{\ss}schedl}}]{Hacar_2016}
{Hacar}, A., {Alves}, J., {Forbrich}, J., {et~al.} 2016, \aap, 589, A80,
  \dodoi{10.1051/0004-6361/201527805}

\bibitem[{{Hernquist}(1990)}]{1990ApJ...356..359H}
{Hernquist}, L. 1990, \apj, 356, 359, \dodoi{10.1086/168845}

\bibitem[{{Hunt} {et~al.}(2019){Hunt}, {Bub}, {Bovy}, {Mackereth}, {Trick}, \&
  {Kawata}}]{2019MNRAS.490.1026H}
{Hunt}, J. A.~S., {Bub}, M.~W., {Bovy}, J., {et~al.} 2019, \mnras, 490, 1026,
  \dodoi{10.1093/mnras/stz2667}

\bibitem[{Hunter(2007)}]{Hunter:2007}
Hunter, J.~D. 2007, Computing in Science \& Engineering, 9, 90,
  \dodoi{10.1109/MCSE.2007.55}

\bibitem[{{Kim} \& {Ostriker}(2018)}]{Kim_2018}
{Kim}, C.-G., \& {Ostriker}, E.~C. 2018, \apj, 853, 173,
  \dodoi{10.3847/1538-4357/aaa5ff}

\bibitem[{{Lada} \& {Lada}(2003)}]{Lada_2003}
{Lada}, C.~J., \& {Lada}, E.~A. 2003, \araa, 41, 57,
  \dodoi{10.1146/annurev.astro.41.011802.094844}

\bibitem[{{Laporte} {et~al.}(2019){Laporte}, {Minchev}, {Johnston}, \&
  {G{\'o}mez}}]{Laporte_2019}
{Laporte}, C. F.~P., {Minchev}, I., {Johnston}, K.~V., \& {G{\'o}mez}, F.~A.
  2019, \mnras, 485, 3134, \dodoi{10.1093/mnras/stz583}

\bibitem[{{L{\'o}pez-Corredoira} {et~al.}(2002){L{\'o}pez-Corredoira},
  {Betancort-Rijo}, \& {Beckman}}]{Lopez_2002}
{L{\'o}pez-Corredoira}, M., {Betancort-Rijo}, J., \& {Beckman}, J.~E. 2002,
  \aap, 386, 169, \dodoi{10.1051/0004-6361:20020229}

\bibitem[{{Matthews} \& {Uson}(2008)}]{Matthews_2008}
{Matthews}, L.~D., \& {Uson}, J.~M. 2008, \apj, 688, 237,
  \dodoi{10.1086/592086}

\bibitem[{{Miyamoto} \& {Nagai}(1975)}]{1975PASJ...27..533M}
{Miyamoto}, M., \& {Nagai}, R. 1975, \pasj, 27, 533

\bibitem[{{Myeong} {et~al.}(2018){Myeong}, {Evans}, {Belokurov}, {Sanders}, \&
  {Koposov}}]{2018ApJ...856L..26M}
{Myeong}, G.~C., {Evans}, N.~W., {Belokurov}, V., {Sanders}, J.~L., \&
  {Koposov}, S.~E. 2018, \apjl, 856, L26, \dodoi{10.3847/2041-8213/aab613}

\bibitem[{{Narayan} {et~al.}(2020){Narayan}, {Dettmar}, \&
  {Saha}}]{Narayan_2020}
{Narayan}, C.~A., {Dettmar}, R.-J., \& {Saha}, K. 2020, \mnras, 495, 3705,
  \dodoi{10.1093/mnras/staa1400}

\bibitem[{{Navarro} {et~al.}(1997){Navarro}, {Frenk}, \&
  {White}}]{1997ApJ...490..493N}
{Navarro}, J.~F., {Frenk}, C.~S., \& {White}, S. D.~M. 1997, \apj, 490, 493,
  \dodoi{10.1086/304888}

\bibitem[{{Nelson} \& {Tremaine}(1995)}]{Nelson_1995}
{Nelson}, R.~W., \& {Tremaine}, S. 1995, \mnras, 275, 897,
  \dodoi{10.1093/mnras/275.4.897}

\bibitem[{{Ness} {et~al.}(2019){Ness}, {Johnston}, {Blancato}, {Rix}, {Beane},
  {Bird}, \& {Hawkins}}]{Ness_2019}
{Ness}, M.~K., {Johnston}, K.~V., {Blancato}, K., {et~al.} 2019, \apj, 883,
  177, \dodoi{10.3847/1538-4357/ab3e3c}

\bibitem[{{Pantaleoni Gonz{\'a}lez} {et~al.}(2021){Pantaleoni Gonz{\'a}lez},
  {Ma{\'\i}z Apell{\'a}niz}, {Barb{\'a}}, \& {Reed}}]{Pantaleoni_2021}
{Pantaleoni Gonz{\'a}lez}, M., {Ma{\'\i}z Apell{\'a}niz}, J., {Barb{\'a}},
  R.~H., \& {Reed}, B.~C. 2021, \mnras, 504, 2968,
  \dodoi{10.1093/mnras/stab688}

\bibitem[{Pedregosa {et~al.}(2011)Pedregosa, Varoquaux, Gramfort, Michel,
  Thirion, Grisel, Blondel, Prettenhofer, Weiss, Dubourg, Vanderplas, Passos,
  Cournapeau, Brucher, Perrot, \& Duchesnay}]{scikit-learn}
Pedregosa, F., Varoquaux, G., Gramfort, A., {et~al.} 2011, Journal of Machine
  Learning Research, 12, 2825

\bibitem[{Price-Whelan(2018)}]{Price-Whelan_2018}
Price-Whelan, A. 2018, \dodoi{10.5281/zenodo.1228136}

\bibitem[{Price-Whelan(2017)}]{Price-Whelan_2017}
Price-Whelan, A.~M. 2017, Journal of Open Source Software, 2, 388,
  \dodoi{10.21105/joss.00388}

\bibitem[{{Reid} {et~al.}(2019){Reid}, {Menten}, {Brunthaler}, {Zheng}, {Dame},
  {Xu}, {Li}, {Sakai}, {Wu}, {Immer}, {Zhang}, {Sanna}, {Moscadelli}, {Rygl},
  {Bartkiewicz}, {Hu}, {Quiroga-Nu{\~n}ez}, \& {van Langevelde}}]{Reid_2019}
{Reid}, M.~J., {Menten}, K.~M., {Brunthaler}, A., {et~al.} 2019, \apj, 885,
  131, \dodoi{10.3847/1538-4357/ab4a11}

\bibitem[{{Sanders} \& {Binney}(2014)}]{2014MNRAS.441.3284S}
{Sanders}, J.~L., \& {Binney}, J. 2014, \mnras, 441, 3284,
  \dodoi{10.1093/mnras/stu796}

\bibitem[{{Sanders} \& {Binney}(2016{\natexlab{a}})}]{Sanders_2016}
---. 2016{\natexlab{a}}, \mnras, 457, 2107, \dodoi{10.1093/mnras/stw106}

\bibitem[{{Sanders} \& {Binney}(2016{\natexlab{b}})}]{2016MNRAS.457.2107S}
---. 2016{\natexlab{b}}, \mnras, 457, 2107, \dodoi{10.1093/mnras/stw106}

\bibitem[{{Speagle}(2020)}]{Speagle_2020}
{Speagle}, J.~S. 2020, \mnras, 493, 3132, \dodoi{10.1093/mnras/staa278}

\bibitem[{{Thulasidharan} {et~al.}(2021){Thulasidharan}, {D'Onghia}, {Poggio},
  {Drimmel}, {Gallagher}, {Swiggum}, {Benjamin}, \& {Alves}}]{Thula_2021}
{Thulasidharan}, L., {D'Onghia}, E., {Poggio}, E., {et~al.} 2021, arXiv
  e-prints, arXiv:2112.08390.
\newblock \doarXiv{2112.08390}

\bibitem[{{Trick} {et~al.}(2019){Trick}, {Coronado}, \& {Rix}}]{Trick_2019}
{Trick}, W.~H., {Coronado}, J., \& {Rix}, H.-W. 2019, \mnras, 484, 3291,
  \dodoi{10.1093/mnras/stz209}

\bibitem[{{Trick} {et~al.}(2021){Trick}, {Fragkoudi}, {Hunt}, {Mackereth}, \&
  {White}}]{Trick_2021}
{Trick}, W.~H., {Fragkoudi}, F., {Hunt}, J. A.~S., {Mackereth}, J.~T., \&
  {White}, S. D.~M. 2021, \mnras, 500, 2645, \dodoi{10.1093/mnras/staa3317}

\bibitem[{Widrow {et~al.}(2014)Widrow, Barber, Chequers, \&
  Cheng}]{Widrow_2014}
Widrow, L.~M., Barber, J., Chequers, M.~H., \& Cheng, E. 2014, Monthly Notices
  of the Royal Astronomical Society, 440, 1971, \dodoi{10.1093/mnras/stu396}

\bibitem[{{Widrow} {et~al.}(2012){Widrow}, {Gardner}, {Yanny}, {Dodelson}, \&
  {Chen}}]{Widrow_2012}
{Widrow}, L.~M., {Gardner}, S., {Yanny}, B., {Dodelson}, S., \& {Chen}, H.-Y.
  2012, \apjl, 750, L41, \dodoi{10.1088/2041-8205/750/2/L41}

\bibitem[{{Zari} {et~al.}(2018){Zari}, {Hashemi}, {Brown}, {Jardine}, \& {de
  Zeeuw}}]{Zari_2018}
{Zari}, E., {Hashemi}, H., {Brown}, A.~G.~A., {Jardine}, K., \& {de Zeeuw},
  P.~T. 2018, \aap, 620, A172, \dodoi{10.1051/0004-6361/201834150}

\bibitem[{{Zucker} {et~al.}(2020){Zucker}, {Speagle}, {Schlafly}, {Green},
  {Finkbeiner}, {Goodman}, \& {Alves}}]{Zucker_2020}
{Zucker}, C., {Speagle}, J.~S., {Schlafly}, E.~F., {et~al.} 2020, \aap, 633,
  A51, \dodoi{10.1051/0004-6361/201936145}

\bibitem[{{Zucker \& Speagle} {et~al.}(2019){Zucker \& Speagle}, {Schlafly},
  {Green}, {Finkbeiner}, {Goodman}, \& {Alves}}]{Zucker_Speagle_2019}
{Zucker \& Speagle}, C. . J.~S., {Schlafly}, E.~F., {Green}, G.~M., {et~al.}
  2019, \apj, 879, 125, \dodoi{10.3847/1538-4357/ab2388}

\end{thebibliography}
\bibliographystyle{aasjournal}



\end{document}